\documentclass[a4paper,fleqn]{cas-dc}

\usepackage{array}
\usepackage{tcolorbox}
\usepackage{dirtytalk}
\usepackage{fontawesome}
\usepackage{pifont}
\usepackage{hyperref}
\usepackage{color, colortbl}
\usepackage{makecell}
\usepackage{balance}
\definecolor{maroon}{rgb}{0.87,0.68,0.32}
\usepackage{ragged2e}
\usepackage{boldline}
\usepackage{xcolor}
\usepackage[T1]{fontenc}
\usepackage{makeidx}
\usepackage[utf8]{inputenc}

\newenvironment{myquote}%
  {\list{}{\leftmargin=0.3in\rightmargin=0.01in}\item[]}%
  {\endlist}

\usepackage[numbers]{natbib}

\definecolor{anti-flashwhite}{rgb}{0.95, 0.95, 0.96}
\definecolor{beige}{rgb}{0.96, 0.96, 0.86}
\definecolor{floralwhite}{rgb}{1.0, 0.98, 0.94}
\definecolor{gainsboro}{rgb}{0.86, 0.86, 0.86}
\definecolor{ghostwhite}{rgb}{0.97, 0.97, 1.0}
\definecolor{honeydew}{rgb}{0.94, 1.0, 0.94}
\definecolor{isabelline}{rgb}{0.96, 0.94, 0.93}
\definecolor{ivory}{rgb}{1.0, 1.0, 0.94}
\definecolor{magnolia}{rgb}{0.97, 0.96, 1.0}
\definecolor{mintcream}{rgb}{0.96, 1.0, 0.98}
\definecolor{pearl}{rgb}{0.94, 0.92, 0.84}
\definecolor{whitesmoke}{rgb}{0.90, 0.90, 0.90}

\def\tsc#1{\csdef{#1}{\textsc{\lowercase{#1}}\xspace}}
\tsc{WGM}
\tsc{QE}
\tsc{EP}
\tsc{PMS}
\tsc{BEC}
\tsc{DE}

\begin{document}


\let\WriteBookmarks\relax
\def\floatpagepagefraction{1}
\def\textpagefraction{.001}
\shorttitle{}
\shortauthors{Ali Rezaei Nasab et~al.}

\title [mode = title]{An Empirical Study of Security Practices for Microservices Systems}

\author[1]{Rezaei Nasab Ali}
\ead{rezaei.ali.nasab@gmail.com}
\address[1]{School of Computer Science, Wuhan University, 430072 Wuhan, China}

\author[2]{Shahin Mojtaba}
\ead{mojtaba.shahin@rmit.edu.au}
\address[2]{School of Computing Technologies, RMIT University, 3000 Melbourne, Australia}

\author[1]{Hoseyni Raviz {Seyed Ali}}
\ead{s.ali.hoseyni@gmail.com}

\author[1]{Liang Peng}
\cormark[1]
\ead{liangp@whu.edu.cn}



\author[3]{Mashmool Amir}
\ead{5245307@studenti.unige.it}
\address[3]{Department of Computer Science, Bioengineering, Robotics and System Engineering, University of Genoa, 16126 Genoa, Italy }

\author[4]{Lenarduzzi Valentina}
\ead{valentina.lenarduzzi@oulu.fi}
\address[4]{Faculty of Information Technology and Electrical Engineering, University of Oulu, FI-90014 Oulu, Finland}

\cortext[cor1]{Corresponding author}


\begin{abstract}
Despite the numerous benefits of microservices systems, security has been a critical issue in such systems. Several factors explain this difficulty, including a knowledge gap among microservices practitioners on properly securing a microservices system. To (partially) bridge this gap, we conducted an empirical study. We first manually analyzed 861 microservices security points, including 567 issues, 9 documents, and 3 wiki pages from 10 GitHub open-source microservices systems and 306 Stack Overflow posts concerning security in microservices systems. In this study, a microservices security point is referred to as ``a GitHub issue, a Stack Overflow post, a document, or a wiki page that entails 5 or more microservices security paragraphs''. Our analysis led to a catalog of 28 microservices security practices. We then ran a survey with 74 microservices practitioners to evaluate the usefulness of these 28 practices. Our findings demonstrate that the survey respondents affirmed the usefulness of the 28 practices. We believe that the catalog of microservices security practices can serve as a valuable resource for microservices practitioners to more effectively address security issues in microservices systems. It can also inform the research community of the required or less explored areas to develop microservices-specific security practices and tools.

\end{abstract}


\maketitle

\section{\textbf{Introduction}}  \label{introduction}

Over the past few years, the Microservices Architecture (MSA) style has been popularly and widely used in the software industry. The MSA style aims to decompose a software application into or build it consisting of a set of microservices (i.e., small business-driven services) that can be implemented, tested, and deployed independently \cite{dragoni2017microservices, fowler2014microservices}. Another benefit of this style is that it allows software development organizations to use the best-fit programming language and technology (e.g., database) to implement each microservice. Several other merits of microservices systems (software systems that employ the MSA style) are pointed out in the literature, such as scalability, modularity, and fault-tolerance \cite{dragoni2017microservices, jamshidi2018microservices}.

Since the advent of the MSA style, securing microservices systems has been a challenge for software practitioners and organizations \cite{di2019architecting, waseem2021nature, rezaei2021automated, pereira2021security, Soldani2018, yarygina2018overcoming, hannousse2021securing}. The potential security challenges associated with microservices systems may compel software organizations to revisit their decision to adopt or migrate to microservices \cite{Mendon9520758, lenarduzzi2020does}. The difficulty in securing microservices systems lies in several factors: (i) Tools and technologies that microservices use or rely on are prone to several security weaknesses and vulnerabilities; (ii) There is a knowledge gap among practitioners and organizations on securing microservices systems as the MSA style is an emerging and evolving architecture style \cite{ghofrani2018challenges, zimmermann2017microservices, Pereira2019SecMec, nadareishvili2016microservice}; (iii) The distributed nature and characteristics of microservices systems make security harder than monolithic systems. For example, it is more difficult to guarantee security in such systems than monolithic systems as hundreds of microservices might be simultaneously running in production. 
Some works have been conducted on security in microservices systems (e.g., \cite{rezaei2021automated, yarygina2018overcoming, pahl2018graph}), and recent review studies have called for more studies on security in microservices systems \cite{dragoni2017microservices, waseem2020systematic, Soldani2018}.


While these works are valuable for security in microservices systems, documented knowledge and guidelines on how software practitioners design and implement secure microservices systems are scarce (if any) \cite{rezaei2021automated}. It is argued that software practitioners are keen to learn and apply the practices and decisions that their peers used or are using during their development process \cite{moore2009crossing, mahdavi2021software, zimmermann2017microservices}. Software practitioners are also interested in learning bad practices adopted by their peers to avoid repeating mistakes. Hezavei et al. \cite{mahdavi2021software} defined a software development practice as “\textit{an activity or step carried out to achieve a goal during the development process}”. To the best of our knowledge, no systematic attempt has been made to develop and document a catalogue of best practices used by software practitioners that enable the development of secure microservices systems. Hence, the ultimate goal of this study is to empirically collect and document microservices-specific security practices. At the same time, it is also important to understand how practitioners perceive the usefulness of these security best practices. To this end, we define the following two research questions (RQs).

 \textit{\textbf{RQ1. What are the security practices to secure microservices systems?}}
 
 \textit{\textbf{RQ2. To what extent do practitioners consider the identified microservices security practices useful?}}

To answer these two RQs, we conducted an empirical study that identified and validated 28 security practices for microservices systems. We first collected and manually analyzed 861 microservices security points, including 567 issues, 9 documents, and 3 wiki pages from 10 GitHub open-source microservices systems and 306 Stack Overflow posts concerning security in microservices systems (see Section \ref{sec:miningsecuritypractices}). A microservices security point is defined as “a GitHub issue, a Stack Overflow post, a document, or a wiki page that entails 5 or more microservices security paragraphs”. This manual analysis led to a catalog of 28 microservices security practices. The security practices identified from GitHub and Stack Overflow are based on the authors’ analysis, which might be subjective and unreliable. Consequently, we ran an industrial survey completed by 74 practitioners to seek their perceptions about the usefulness of the identified security practices (see Section \ref{sec:validationsurvey}).

The key contributions of this paper can be summarized as follow:
\begin{itemize}
    \item Identification of 28 security practices in 6 categories for microservices systems;
    \item Validation of the usefulness of these security practices from 74 microservices practitioners;
    \item Providing an online replication package  of the data used in this study for researchers and practitioners to replicate and validate the findings \cite{onlinedataset};
    \item A set of actionable recommendations for microservices practitioners and researchers.
\end{itemize}

The rest of the paper is organized as follows: Section \ref{backgroundandrelatedwork} provides the background on microservices systems and their security and summarizes the related work. Section \ref{methodology} details our research methodology. The findings are presented in Section \ref{Findings}, followed by a set of recommendations for practitioners and researchers in Section \ref{recommendations}. Section \ref{Threats-to-Validity} elaborates on the threats of our study. Section \ref{ConclusionsandFutureWork} concludes our work and provides future work directions.

\section{Background and Related Work}\label{backgroundandrelatedwork}

\subsection{Background}

\subsubsection{Microservices Systems}\label{microservices-systems}

Despite the MSA style being originated from Service-Oriented Architecture (SOA), they have some significant differences \cite{dragoni2017microservices, fowler2014microservices}. Both include (small) services with dedicated responsibilities, while services in the MSA style are independent and autonomous and communicate through different lightweight mechanisms, services in SOA are not full-stack and fully autonomous \cite{Pahl2018}.

The development, test, and deployment of each microservice can be done independently by a different development team using divergent technologies and programming languages. The unique characteristics of the MSA style allow microservices to scale independently from each other \cite{jamshidi2018microservices}. Furthermore, the MSA style enables the development teams to use the hardware that adequately meets their needs to deploy each microservice. As microservices are small and independent, their maintenance and making them fault-tolerant would be much easier. It is because the failure of one service will not lead to the entire system down, which may occur in monoliths~\cite{fowler2014microservices}.

The research and industry communities have investigated several aspects of microservices systems. One of the most prevailing investigated and demanding aspects is how to migrate a legacy/monolithic application to microservices \cite{di2019architecting, waseem2020systematic}. For example, Balalaie et al. \cite{balalaie2016microservices} identified 15 patterns for this purpose, and Auer et al. \cite{auer2021monolithic} developed an assessment framework to help software organizations specify and measure possible advantages and difficulties of migration to microservices. However, the migration to or adoption of microservices can be associated with many serious challenges and issues, which need careful consideration \cite{Soldani2018, TaibiIEEE2017}. These issues and challenges are enormous, such as the cost overhead due to the migration, the higher complexity of the system due to the increased amount and variety of components integrated into the system, and security issues \cite{Mendon9520758, lenarduzzi2020does}.

Some researchers also looked at approaches and tools for microservices systems. Cinque et al. \cite{cinque2019microservices} developed an approach for monitoring microservices, and Heorhiadi et al. \cite{heorhiadi2016gremlin} introduced a framework, \textit{Gremlin}, to test the \say{failure-handling capabilities} of microservices. Other researchers empirically investigated how microservices systems are designed, implemented, tested, and monitored in the software industry (e.g., \cite{waseem2021design, bogner2019microservices}).

\subsubsection{Security in Microservices Systems}\label{security-in-microservices-systems}
Besides the various advantages brought by microservices, security often becomes an issue during their deployment. Similar to other types of systems (e.g., monolithic systems), improved security in microservices systems may be accomplished in different ways, such as by applying a secure development methodology \cite{matulevivcius2017fundamentals, washizaki2021systematic, uzunov2018assessing}. A monolithic system is a single system, and usually, a single application server needs to be secured, while in microservices systems, each microservice represents a possible attack surface~\cite{pereira2021security}. In monolithic systems, communications between different components happen locally, with local calls, while in microservices systems, the communication happens through the network, creating another possible attack surface. In the MSA style, a compromised microservice can send malicious requests to other microservices. 

Another aspect to consider is the authorization between services since not all the services might be authorized to connect to other services. Different systems, such as Kubernetes\footnote{\url{https://kubernetes.io}} or Istio\footnote{\url{https://istio.io}} provide inter-service authorization mechanisms. However, these mechanisms need to be developed, and new distributed access rules need to be defined separately for each service. The authorization mechanisms should try to reduce the privileges between services as much as possible instead of opening granting full access to all the services~\cite{pereira2021security}.

When dealing with authentications, distributed authentication mechanisms need to be considered. As an example, developers need to decide how to handle authentication, if using an authentication server, or independent authentication systems in different microservices \cite{pereira2021security}. Moreover, developers need to discriminate how to update the authentication mechanisms every time when new services or new users are included in the system.

The security and authentication issues need to be carefully designed by the architects, who are designing a specific system. Microservices systems also need to consider possible vulnerabilities due to the usage of public container images that might be potentially infected \cite{scott2018MSACont}. An attempt to reduce this issue is provided by the \say{Moving Target Defenses}, which proposes to modify component images to create uncertainty for attackers~\cite{Torkura2018}.


Moreover, when migrating from a monolithic system to microservices, companies need to keep the monolithic system and the microservices alive and connected at the same time until the migration is completed and the monolithic system is shut-down~\cite{Soldani2018, TaibiIEEE2017}. This requires creating a secure communication channel between the monolithic system and the microservices system and integrating the security and authentication system adopted for the monolithic system with the new one adopted in the microservices system~\cite{Soldani2018, TaibiIEEE2017}. The main challenges are to identify a suitable approach to: 
\begin{itemize}
    \item Manage and synchronize the authentication of microservices with the monolithic system, so as that end-users will not realize that the system is being migrated; 
    \item Secure the microservices, the communication between the monolithic system and the microservices, but also the communication intra-microservices ;
    \item Understand which type of vulnerabilities are introduced during the  migration.
\end{itemize}

\subsection{Related Work}
Researchers and practitioners may use security techniques, tools, practices, patterns, or secure development methodologies to support the development of secure software systems \cite{matulevivcius2017fundamentals, washizaki2021systematic, uzunov2018assessing}. This section covers the studies that have targeted security in microservices systems.

Several secondary studies have been conducted on microservices systems and on security in microservices. In a recent review study, Waseem et al. \cite{waseem2020systematic} observed that there are a few concrete solutions for addressing security concerns when implementing microservices systems in DevOps. Another systematic mapping review on 46 papes by Hannousse and Yahiouche \cite{hannousse2021securing} revealed that most studies on security in microservices are the solutions proposed in the soft-infrastructure layer. They argued that internal attacks, compared to external attacks, are less explored in the literature. They also indicated more efforts should be allocated on other layers of MSA (e.g., communication and deployment layers) and developing mitigation techniques. Pereira-Vale et al. \cite{pereira2021security} carried out a multivocal literature review on 36 academic literature and 34 grey literature. Their review led to a classification of 15 security mechanisms, in which authentication and authorization are the security mechanisms most reported in the literature. In addition, \say{mitigate/stop attacks} are expressed in about 2/3 of the security mechanisms. 
Ponce et al. \cite{ponce2021smells} focused on security smells and refactorings and collected and analyzed 58 white and grey literature published from 2014 to 2021. They found 10 security smells with their security properties and corresponding refactorings. They also elaborated on how the corresponding refactorings mitigate the effects of the smells. A recent systematic grey literature study by Billawa et al. on 57 microservices-related grey literature sources identified 7 security challenges, including ``\textit{trust between services}'', ``\textit{large attack area}'', ``\textit{testing}'', ``\textit{container management}'', ``\textit{low visibility}'', ``\textit{secret management}'', and ``\textit{polyglot architecture}'' \cite{billawa2022security}. Billawa et al. also found that the practices, such as ``\textit{defense in depth}'', ``\textit{DevSecOps}'', ``\textit{encrypt sensitive data}'', ``\textit{immutable container}'', ``\textit{rate throttling}'', ``\textit{secure-by-design}'', and ``\textit{least privilege}'' can help develop secure microservices systems. 


Waseem et al. \cite{waseem2021nature} empirically investigated the issues reported in 5 open source microservices systems hosted on GitHub. They found that 10.18\% of these issues can be attributed to security. Yarygina and Bagge \cite{yarygina2018overcoming} asserted that security in microservices systems is a multi-faceted problem, and a layered security solution can address it. Hence, they categorized microservices security concerns into 6 layers (hardware, virtualization, cloud, communication, service/application, and orchestration) with solutions to address them (e.g., secure implementation of service discovery and registry components). Richardson \cite{richardson2018microservices} introduced 44 microservices patterns that are categorized into 16 groups (e.g., “data consistency” patterns). Among them, “Access Token” is the only pattern related to security.


To secure IoT microservices, Pahl et al. \cite{pahl2018graph} proposed a graph-based access control module in a network of IoT nodes. This module can monitor the communication of IoT microservices systems to create robust security and mitigate the security holes of such systems. Moreover, Pahl and Donini \cite{pahl2018securing} provided a method based on “X.509 certificates” for authenticating IoT microservices. One of the main goals of this method is to verify the security properties locally through the distributed IoT nodes. Yu et al. \cite{yu2019survey} focused on security issues of microservices-based fog applications because such systems are composed of numerous microservices with complex communications, which is a challenge in terms of security. They reviewed 66 articles and identified 17 security issues (e.g., kernel exploit or DOS attack) regarding the microservices communication categorized in 4 groups: containers issues, data issues, permission issues, and network issues.

By surveying 67 participants, Rezaei Nasab et al. \cite{rezaei2021automated} affirmed that security challenges in microservices systems differ from non-microservices systems. To bridge the security knowledge gap among microservices practitioners \cite{ghofrani2018challenges, zimmermann2017microservices, Pereira2019SecMec, nadareishvili2016microservice}, they developed a set of machine and deep learning approaches to automatically recognize security discussions (including security design decisions, challenges, or solutions) from open source microservices systems. Chondamrongkul et al. \cite{chondamrongkul2020automated} also developed an automated approach using ontology reasoning and model checking techniques to identify security threats of microservices architectures through analyzing security characteristics. 
The identified security threats show how the attack scenarios may happen. Sun et al. \cite{sun2015security} designed an API primitive FlowTap that provides security-as-a-service for microservices-based cloud applications. The proposed technique can monitor and protect the network of such systems from internal and external threats. 

In contrast to the works above, our study presents 28 trusted and ready-to-use security best practices for microservices systems. For example, although practices identified in \cite {billawa2022security} are valuable, they are still general and need concrete solutions to be implemented in microservices system development. Our 28 practices were collected from developer discussions occurring during the development of microservices systems. We further validated the usefulness of these practices by seeking feedback from 74 microservices practitioners. Finally, we articulated the positive and negative sides (if any) of the 28 practices.

\section{Methodology}\label{methodology}

Our goal in this study is to identify and evaluate a catalog of security practices for microservice systems. In this section, we first explain our approach to identifying 28 security practices for microservice systems from GitHub and Stack Overflow (RQ1) in Section \ref{sec:miningsecuritypractices}. We then discuss the design and execution of a survey to evaluate these security practices (RQ2) in Section \ref{sec:validationsurvey}.
\begin{figure*}
    \centering
    \includegraphics[scale=.89]{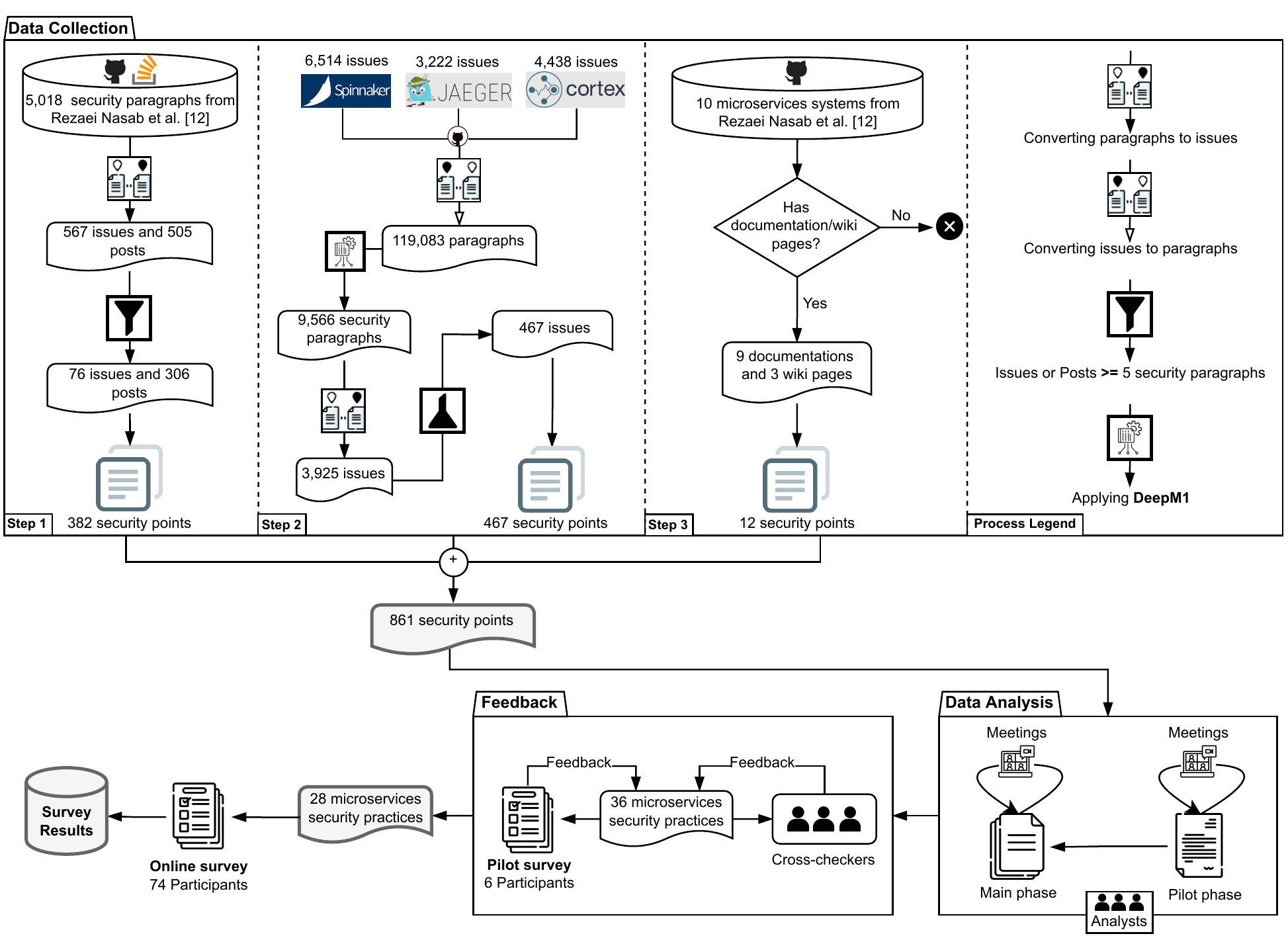}
    \caption{An overview of extracting microservices security practices}
    \label{fig:overview}
\end{figure*}

\subsection{Mining Security Practices (RQ1)} \label{sec:miningsecuritypractices}

In this section, we first describe how we collected data related to microservices security from GitHub and Stack Overflow (see Section~\ref{subsec:datacollection}), and then how the collected data was analyzed to identify microservices security practices (see Section \ref{subsec:identificationofsecuritypractices}).


\subsubsection{Data Collection}\label{subsec:datacollection}
 GitHub issues and Stack Overflow posts (including Stack Exchange) are valuable sources for identifying development practices (e.g., architectural practices \cite{bi2021mining, malavolta2021mining} and security practices \cite{meng2018secure}). Hence, we focused on GitHub and Stack Overflow to collect data related to microservices security.
 
\textbf{Step 1.} In our previous work \cite{rezaei2021automated}, we manually created a dataset of 5,018 security paragraphs collected from 567 GitHub issues in 7 open-source microservices systems and 505 Stack Overflow posts. These security paragraphs include \textit{“design decisions, challenges, or solutions relating to security in microservices systems”} \cite{rezaei2021automated}. The architectural style of these 7 open-source systems (i.e., goa, eShopOnContainers, microservices-demo, scalecube-services, moleculer, deep-framework, and light-4j shown in Table \ref{MSAprojects}) was determined as the microservices architecture style by enquiring the core contributors of those projects through an online survey. The core contributors are defined as \textit{“the top 3 contributors who have the most commits in a project”} \cite{rezaei2021automated}. Our previous work \cite{rezaei2021automated} aimed to develop ML/DL-based approaches to automatically differentiate security paragraphs from non-security paragraphs in GitHub issues and Stack Overflow posts concerning security in microservices. In contrast, this work aims to identify security best practices for microservices systems from GitHub and Stack Overflow and then evaluate them with software practitioners.

Hence, the 5,018 security paragraphs collected from our previous work could be a good source to identify security practices. Despite this, the security paragraphs are short (i.e., 2-3 sentences), and there might be the chance of losing the design context by reading the security paragraphs individually. Our strategy to (partially) mitigate this challenge was to read the entire GitHub issues or Stack Overflow posts that entail equal or more than 5 such security paragraphs. Our decision to set the threshold to 5 security paragraphs was based on our experience in the previous work \cite{rezaei2021automated} as issues and posts with 5 or more security paragraphs are more suitable for identifying security practices.
As shown in Figure \ref{fig:overview}, this process led to 76 candidate GitHub issues and 306 candidate Stack Overflow posts, which are called security points. The creation date of the 306 candidate Stack Overflow posts ranges from August 2014 to July 2020.

\textbf{Step 2.} Apart from the 7 open-source microservices systems used in our previous work \cite{rezaei2021automated}, we found 3 larger open-source microservices systems (spinnaker with 6,514 issues, jaeger with 3,222 issues, and cortex with 4,438 issues shown in Table \ref{MSAprojects}) on GitHub. These 3 new systems also employ the MSA style. We used the same approach in \cite{rezaei2021automated} (i.e., enquiring the core contributors) to identify these new microservices systems. 

We applied DeepM1 (i.e., the best performing ML/DL approach developed in \cite{rezaei2021automated}) to identify security paragraphs from the GitHub issues of these 3 new projects. Since the unit of analysis in the DeepM1 approach \cite{rezaei2021automated} was a paragraph, we converted the GitHub issues of these 3 projects to paragraphs using their HTML tag $<$p$>$. The pre-processing used in \cite{rezaei2021automated} was also used for these paragraphs. Note that DeepM1 with a recall of 84.25\% and a precision of 86.73\% works at the paragraph level. This process led to identifying 9,566 security paragraphs from the GitHub issues of these 3 new systems, which include 5,931 security paragraphs from spinnaker project, 1,629 security paragraphs from jaeger project, and 2,006 security paragraphs from cortex project (see Figure \ref{fig:overview}). Similar to the approach used in \textbf{Step 1}, we only selected the GitHub issues of these 3 projects that entail equal or more than 5 such security paragraphs. This resulted in a collection of 467 security points from spinnaker, cortex, and jaeger projects (see Figure \ref{fig:overview}).

\textbf{Step 3.} Among all systems discussed in \textbf{Step 1} and \textbf{Step 2}, 9 systems (i.e., eShopOnContainers, spinnaker, cortex, jaeger, goa, moleculer, deep-framework, microservices-demo, and light-4j) have documentation with 5 or more security paragraphs. Further, 3 systems, including eShopOnContainers, scalecube-services, and light-4j, have a wiki page with 5 or more security paragraphs. We considered these 9 pieces of documentation and 3 wiki pages as security points. As shown in Figure \ref{fig:overview}, our dataset includes 861 security points, which were manually analyzed to identify security practices (see Section \ref{subsec:identificationofsecuritypractices}).

\subsubsection{Identification of Security Practices}\label{subsec:identificationofsecuritypractices}
Identifying and extracting microservices security practices from the 861 microservices security points include 3 phases.

\textbf{Pilot Phase.} We randomly selected 20 security points from the 861 security points and asked 3 analysts (3 authors) to analyze them independently. The goal was to get familiar with data and understand what sort of security points should be considered a security practice. Each analyst applied the open coding and constant comparison techniques from Grounded Theory \cite{glaser1968discovery} to extract security practices. They then held a meeting to check the similarities and dissimilarities of the extracted security practices and resolve any disagreements.

{\renewcommand{\arraystretch}{1}
\begin{table*}[t]
\caption{Demographic questions of our online survey}\label{demo-questions}
\resizebox{\textwidth}{!}{
\begin{tabular}{lll}
\Xhline{3\arrayrulewidth}
\textbf{Demographic Questions}                                                            & \textbf{Question Type}   & \textbf{Example Answers}                                                        \\ \Xhline{2.3\arrayrulewidth}
“How many years have you been involved in software development?”                   & Multiple choice & 0 \textless year \textless{}= 2  \\
“What is your main role in software development?”                                  & Multiple choice & Developer, Architect, Tester                                     \\
“How many years have you been involved in microservices system development?”       & Multiple choice & 0 \textless year \textless{}= 1  \\
“How many years of experience do you have with security in microservices systems?” & Multiple choice & 0 \textless year \textless{}= 1  \\
“How large is your organization?”                                                  & Multiple choice & 20 \textless{}= employees \textless{}= 50 \\
“What are the domains of your organization?”                                       & Checkbox     &  Financial, E-commerce                             \\
“Which country do you currently work in?”                                          & Free Text    & Australia                                                  \\ \Xhline{3\arrayrulewidth}
\end{tabular}}
\end{table*}}

\textbf{Main Phase.}  The 3 analysts collaboratively analyzed the remaining 841 security points. 90 security points were allocated to the 3 analysts each week. In other words, each analyst was asked to extract security practices from 30 allocated security points using the open coding and constant comparison techniques \cite{glaser1968discovery}. Further, an Excel file was created and shared with all the analysts. They were requested to maintain the link between an identified practice and its corresponding security point. At the end of each week, the analysts held a meeting to discuss their extracted security practices, identify the duplicate ones, merge, or rephrase them. This process led to the identification of 36 practices, which were grouped into 6 categories based on their topics: \textit{Authorization and Authentication}, \textit{Token and Credentials}, \textit{Internal and External Microservices}, \textit{Microservices Communications}, \textit{Private Microservices}, and \textit{Database and Environments}.

\textbf{Feedback Phase.} In this phase, the other 3 authors reviewed the 36 security practices. They mainly checked the 36 security practices to identify and mitigate possible inconsistencies and ambiguities. This step resulted in merging 4 practices with other microservices security practices and reducing the number of security practices to 32 practices. 

Next, we ran a pilot survey to seek microservices practitioners' feedback on the 32 identified security practices. The pilot survey was designed using Google Forms and completed by 6 microservices practitioners. We asked the practitioners to demonstrate their level of agreement or disagreement with the 32 identified security practices (Likert scale questions rated from “strongly agree = 5” to “strongly disagree = 1”). Each Likert scale question was followed by an optional open-ended question to obtain further feedback. Based on feedback collected from the practitioners and our internal discussions, we reduced the number of security practices from 32 to 28. This reduction was because the practitioners indicated that 2 security practices did not contain enough information to be understood. We also found 2 practices that overlapped and merged them. We further slightly rephrased the wording of some practices (e.g., adding more information to a security practice) to remove any ambiguities. These 28 security practices are distributed into the 6 categories developed in the \textbf{Main Phase} as follows: \textit{Authorization and Authentication} (6 practices), \textit{Token and Credentials} (5 practices), \textit{Internal and External Microservices} (7 practices), \textit{Microservices Communications} (4 practices), \textit{Private Microservices} (2 practices), and \textit{Database and Environments} (4 practices).

\subsection{Validation Survey (RQ2)} \label{sec:validationsurvey}
In this section, we elaborate on the design and execution of a survey (called the validation survey) to evaluate the usefulness of the 28 security practices for microservices systems collected in Section \ref{sec:miningsecuritypractices}.

\subsubsection{Protocol} \label{protocol}
Considering the guidelines proposed by Kitchenham and Pfleege \cite{kitchenham2008personal}, an online survey (i.e., the validation survey) was developed to evaluate the usefulness of the 28 microservices security practices identified in Section \ref{sec:miningsecuritypractices}. The survey was anonymous and hosted on Google Forms. The survey preamble describes the goal of the survey and briefly explains how and from which sources these 28 microservices security practices are identified. The survey includes 45 questions and takes about 25 minutes to complete. The survey questions can be classified into 3 groups.

\textbf{Demographic questions}. We asked 7 questions to get background information about the survey participants (e.g., “\textit{how many years of experience do you have with security in microservices systems?}”). Table \ref{demo-questions} shows these 7 questions. All demographic questions except one were compulsory.

\textbf{Likert scale questions}. The respondents were asked to rate the usefulness of each of the 28 identified security practices using a Likert question. The mandatory Likert scale questions were ranked on a 4-point scale as “\textit{Absolutely Useful}”=4, “\textit{Useful}”=3, “\textit{Not Useful}”=2, and “\textit{Absolutely Not Useful}”=1. We also added the option “\textit{I Don't Know}” to allow practitioners not to respond to the practices when they were unsure about or unclear to them.
  
\textbf{Open-ended questions}. As discussed in Section \ref{subsec:identificationofsecuritypractices}, the 28 security practices are classified into 6 categories. For each category, we asked the participants to provide the reason for their response for one of the practices in that category that they rated “\textit{Absolutely Useful/Useful}” or “\textit{Absolutely Not Useful/Not Useful}”. The participants were also requested to list the practice number. Note that answering these questions was optional. Finally, 2 more optional questions were asked. The respondents were requested to share any feedback about the security practices in microservices systems. The second question was to allow the participants to provide their email addresses if they were interested in the results of our study.

\subsubsection{Participants} \label{participants}

We used the following methods to recruit microservices practitioners.
\\
\ding{202} We collected the publicly available emails of 868 contributors involved in the 10 projects listed in Table \ref{MSAprojects}. We emailed them and asked them to fill up the survey. 
\\
\ding{203} The spinnaker project has a workplace on Slack with many active contributors. The workplace has some Special Interest Group (SIG) channels that focus on different topics (e.g., security). We advertised our survey in the workplace.
\\
\ding{204} The third strategy was to broadly advertise the survey in some microservices groups on social networks like Twitter and LinkedIn. Further, we sent private messages to practitioners who were a member of these groups. 
\\
\ding{205} We asked the invited practitioners to share the survey with their colleagues who had experience in microservices security.

In total, we received 74 valid responses. The initial analysis of the survey responses revealed that 2 responses were invalid. For example, one participant answered all 28 Likert questions as “\textit{I Don't Know}”. Note that we did not calculate the response rate for our survey because of our heterogeneous recruitment process (e.g., the respondents might contribute to one or more projects in Table \ref{MSAprojects}, and at the same time, they might be involved in multiple LinkedIn groups).

{\renewcommand{\arraystretch}{1.3}
\begin{table*}[]
\caption{A list of 10 microservices systems used in this study. Release (Rel.); Contributors (Cont.); Line of Codes (LoC) }\label{MSAprojects}
\centering
\resizebox{\textwidth}{!}{
\begin{tabular}{llllllllllll}
\hline
\rowcolor[HTML]{FFFFFF} \textbf{\#} &
\textbf{Project Name} & \textbf{URL}           & \textbf{Stars} & \textbf{Forks} & \textbf{Issues} & \textbf{Rel.} & \textbf{Cont.} & \textbf{LoC} & \textbf{Languages} & \textbf{Docs} & \textbf{Wiki} \\ \Xhline{2.3\arrayrulewidth}
\rowcolor[HTML]{FFFFFF} 
1 & eShopOnContainers     & \url{https://bit.ly/2X40b4M} & 18.6k          & 7.9k           & 1,752           & 17                & 142                   & 120k         & C\#    &   \ding{52}   &     \ding{52}     \\
\rowcolor[HTML]{EFEFEF} 
2 & jaeger                & \url{https://bit.ly/2YvyJgN} & 14.2k          & 1.7k           & 3,222           & 44                & 221                   & 105k         & Go, Shell       &   \ding{52}   &     \ding{56}    \\
\rowcolor[HTML]{FFFFFF} 
3 & spinnaker             & \url{https://bit.ly/3ndjJyu} & 8k             & 1.1k           & 6,514           & 132               & 116                   & 6k           & Shell, Go      &   \ding{52}   &    \ding{56}      \\
\rowcolor[HTML]{EFEFEF} 
4 & moleculer             & \url{https://bit.ly/2X5DayD} & 4.6k           & 441            & 1,003           & 99                & 92                    & 98k          & Javascript       &   \ding{52}   &  \ding{56}      \\
\rowcolor[HTML]{FFFFFF} 
5 & goa                   & \url{https://bit.ly/38LUYkD} & 4.4k           & 459            & 2,907           & 55                & 88                    & 88k          & Go            &   \ding{52}   &   \ding{56}        \\
\rowcolor[HTML]{EFEFEF} 
6 & cortex                & \url{https://bit.ly/3nbq65x} & 4.3k           & 604            & 4,438           & 50                & 204                   & 1.2m         & Go          &  \ding{52}    &      \ding{56}       \\
\rowcolor[HTML]{FFFFFF} 
7 & light-4j              & \url{https://bit.ly/3zOnhL0} & 3.3k           & 554            & 1,030           & 140               & 34                    & 57k          & Java        &   \ding{52}   &     \ding{52}        \\
\rowcolor[HTML]{EFEFEF} 
8 & microservices-demo    & \url{https://bit.ly/38LtxY2} & 2.8k           & 1.8k           & 875             & 13                & 55                    & 15k          & Python         &  \ding{52}    &   \ding{56}       \\
\rowcolor[HTML]{FFFFFF} 
9 & deep-framework        & \url{https://bit.ly/3ncbgM4} & 538            & 75             & 647             & 22                & 12                    & 956k         & Javascript      &    \ding{52}  &    \ding{56}     \\
\rowcolor[HTML]{EFEFEF} 
10 & scalecube-services    & \url{https://bit.ly/3DTFhGn} & 507            & 79             & 820             & 178               & 21                    & 11k          & Java        &   \ding{56}   &    \ding{52}         \\ \hline
\end{tabular}}
\end{table*}}

\subsubsection{Data Analysis}\label{dataanalysis}

Descriptive statistics were used to study the responses to the closed-ended questions, i.e., demographic and Likert scale questions. We also applied the open coding technique to analyze the responses to the open-ended questions \cite{glaser1968discovery}. Note that we used the answers (if any) to the open-ended questions to clarify why a particular security practice was chosen \say{\textit{Useful/Absolutely Useful}} or \say{\textit{Not Useful/Absolutely Not Useful}} by the respondents.

\section{Findings}\label{Findings}

\begin{figure*}
    \centering
    \includegraphics[scale=.52]{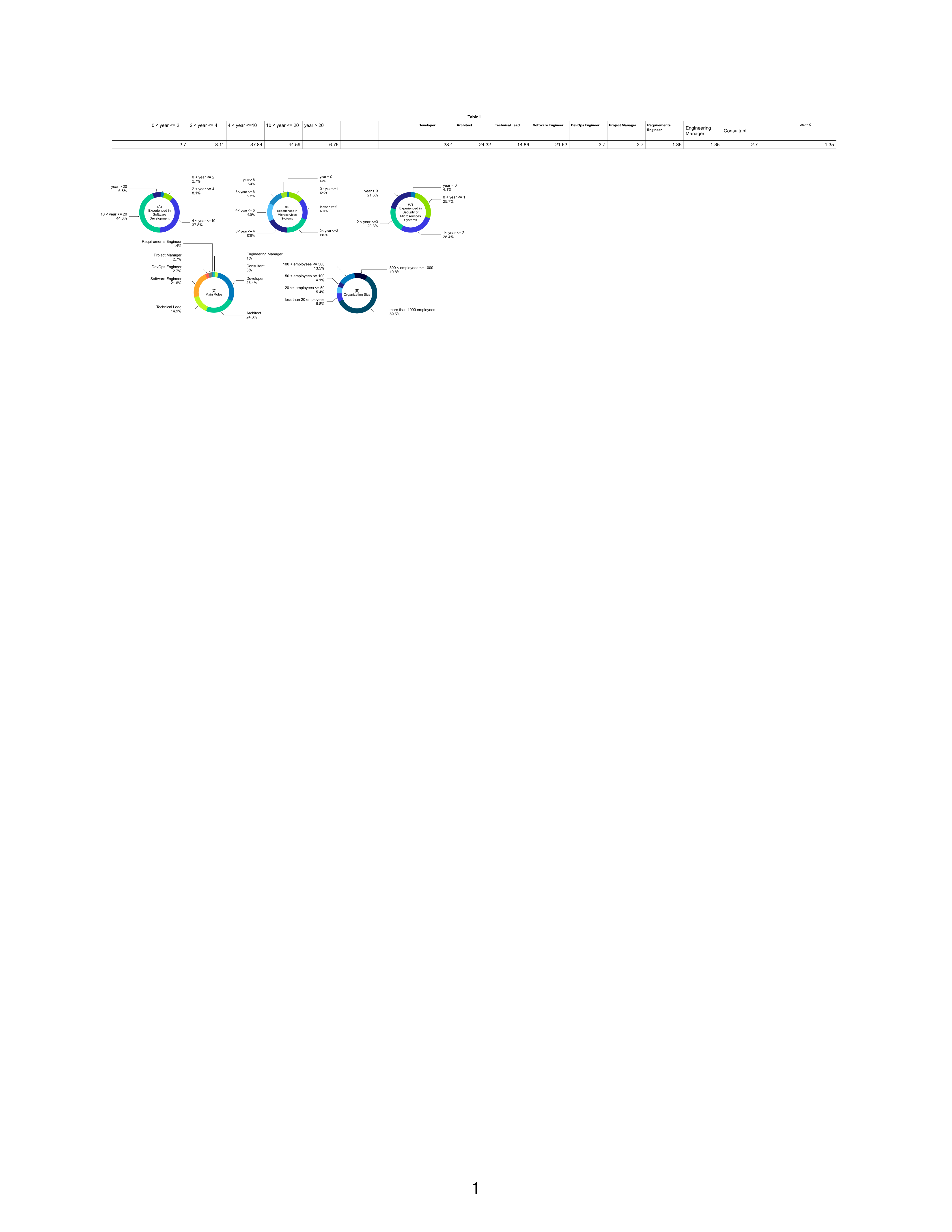}
    \caption{Experience of the participants (n=74) in software development (\textbf{A}), experience of the participants in microservices system development (\textbf{B}), experience of the participants in securing microservices systems (\textbf{C}), main roles of the participants (\textbf{D}), and organization size of the participants (\textbf{E}).}
    \label{fig:demo_5q}
\end{figure*}

\begin{figure}
    \centering
    \includegraphics[scale=.49]{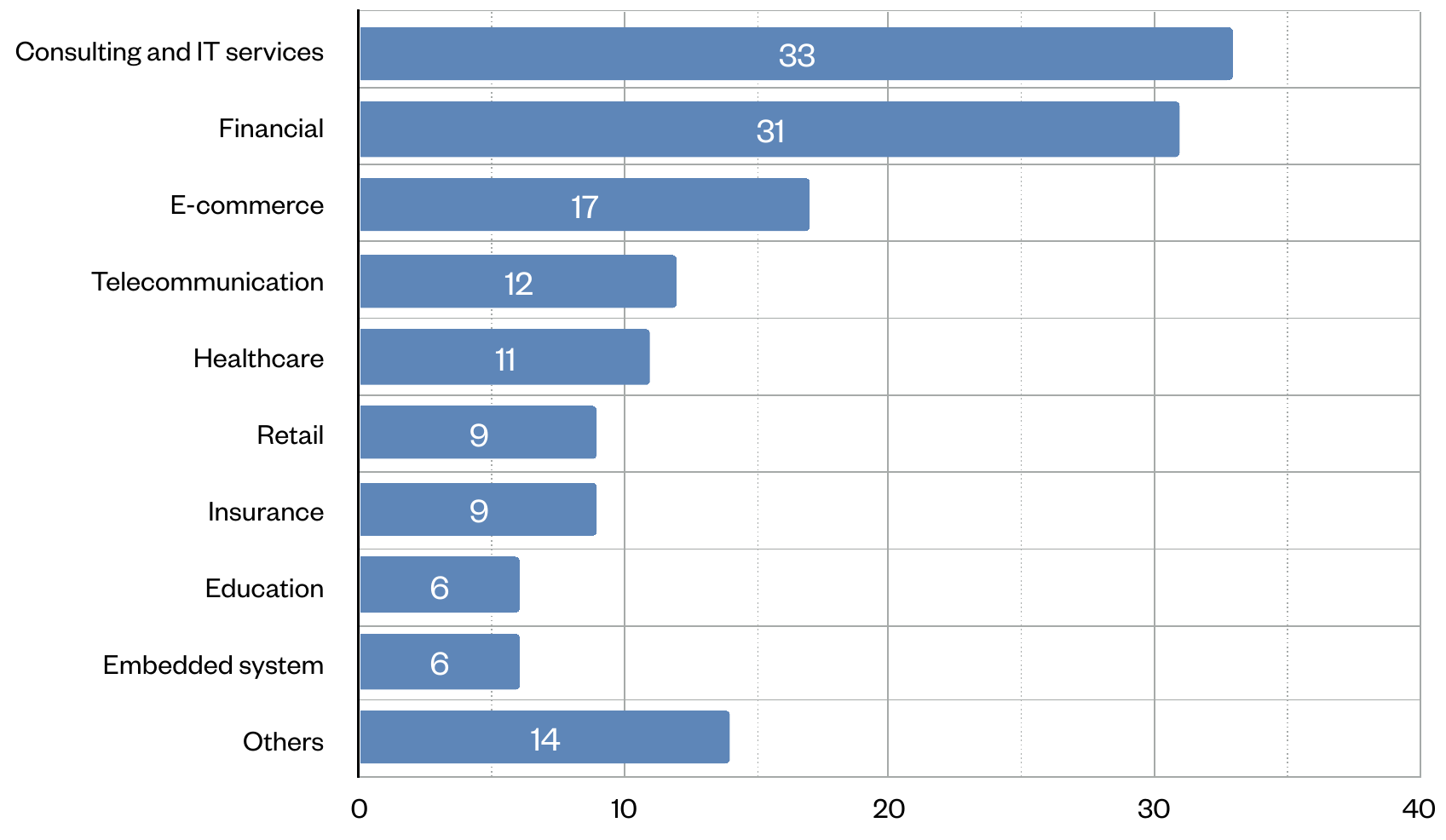}
    \caption{Participants' organization domains (n=74). Note: Participants could select more than one domain}
    \label{fig:demo_domain}
\end{figure}

\begin{figure}
    \centering
    \includegraphics[scale=.52]{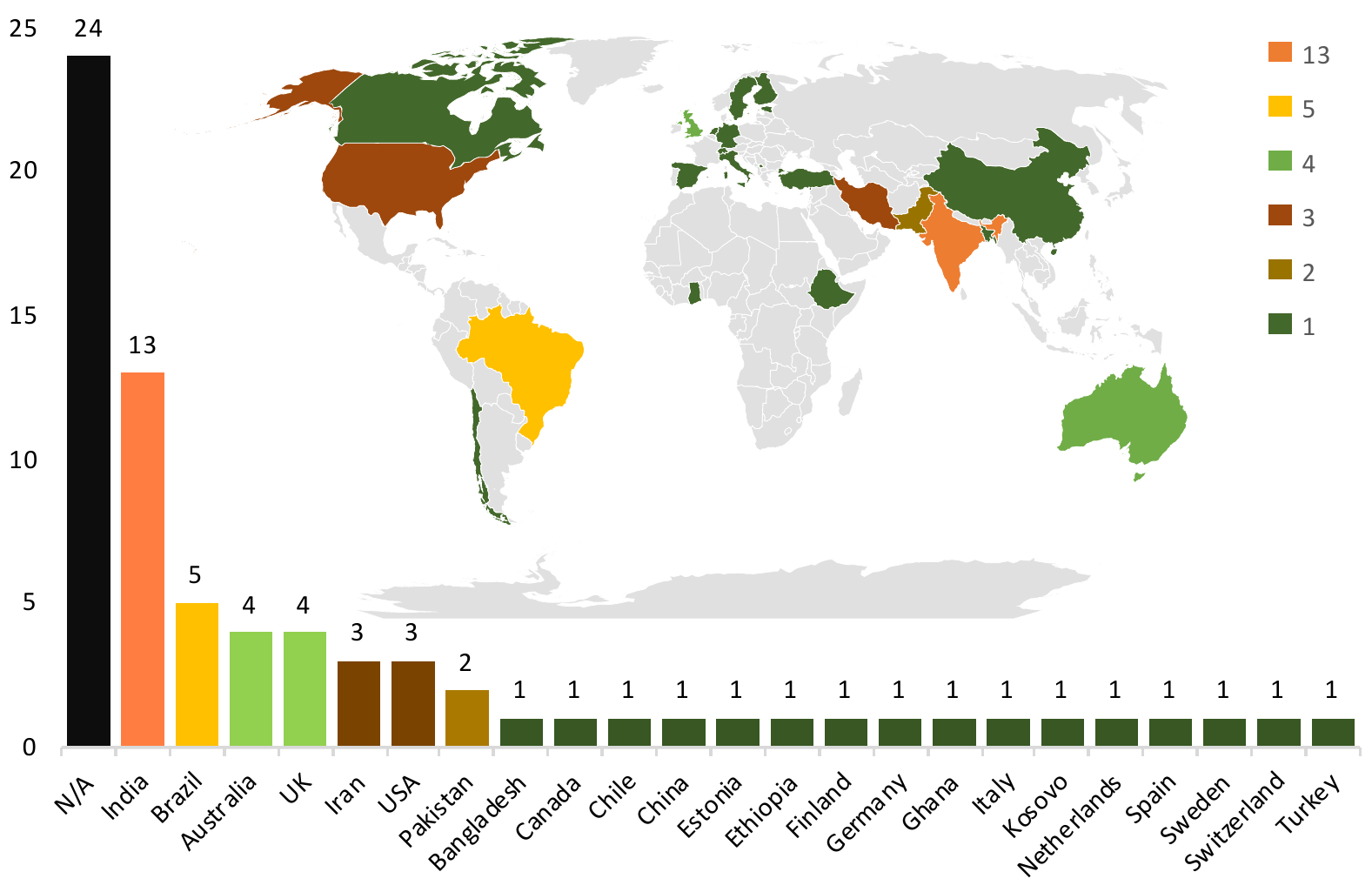}
    \caption{Number of participants (n=74) from 23 countries}
    \label{fig:demo_country}
\end{figure}

\subsection{Demographics}\label{Demographics}
We provide the demographics of the survey respondents.

\textbf{Experience.} Figure \ref{fig:demo_5q} (A) shows that 51.4\% of the participants (n=74) have been involved in software development for at least 10 years. All participants, except for one participant, had at least one year of experience in developing microservices systems, with 50.1\% having more than 3 years of experience (see Figure \ref{fig:demo_5q} (B)). Regarding Figure \ref{fig:demo_5q} (C), more than 70\% of the respondents had one year of experience with securing microservices systems. 25.7\% worked with security in microservices systems in less than one year. The rest (4.1\%) did not have any experience in this regard.

\textbf{Role.} As shown in Figure \ref{fig:demo_5q} (D), the participants mainly worked as Developer (28.4\%, 21 out of 74), Architect (24.3\%, 18 out of 74), Software Engineer (21.6\%, 16 out of 74), and Technical Lead (14.9\%, 11 out of 74).

\textbf{Organization size and domain.} The majority of the participants (83.8\%) came from organizations with more than 100 employees (see Figure \ref{fig:demo_5q} (E)). 59.5\% (44 out of 74 participants) were from organizations with more than 1000 employees. The participants' organization domains are shown in Figure \ref{fig:demo_domain}. The participants were able to choose one or more organization domains in the demographic question. The dominant domains are ``consulting and IT services'' and ``financial'', followed by ``E-commerce'' and ``telecommunications''.

\textbf{Country.} The distribution of participants per country is shown in Figure \ref{fig:demo_country}. Since this question was optional, we only received 50 responses for this question. The 50 participants who indicated their country information came from 23 countries across 6 continents, including Europe (10 countries), Asia (6 countries), North America (2 countries), South America (2 countries), Africa (2 countries), and Oceania (1 country). Most of them were from India, Brazil, and Australia (see Figure \ref{fig:demo_country}).

\begin{figure}
    \centering
    \includegraphics[scale=0.55]{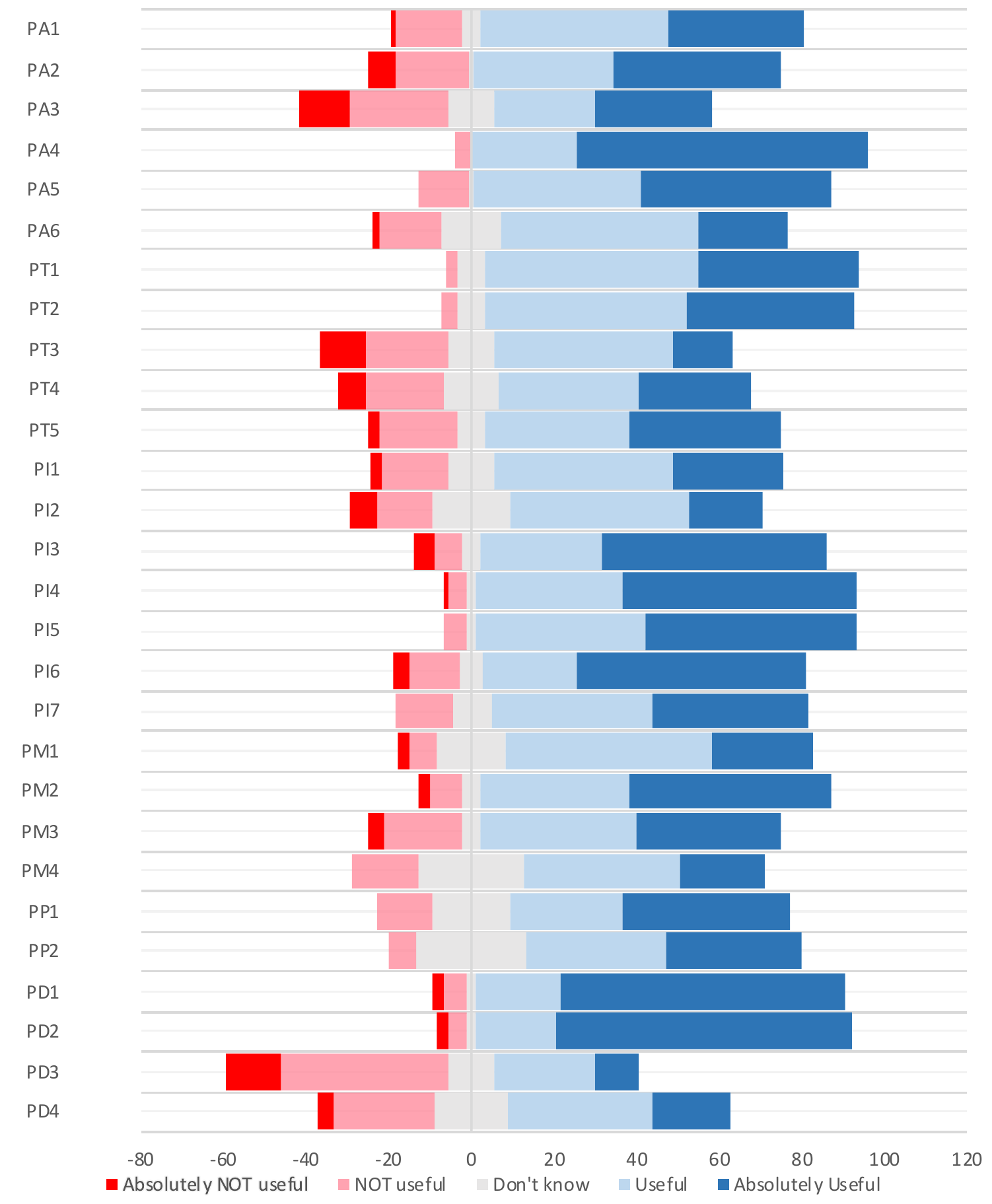}
    \caption{The level of usefulness of the 28 identified security practices for microservices systems from the perspective of 74 practitioners}
    \label{fig:usefulness_practices}
\end{figure}

\subsection{Security Practices}\label{Security-Practices}
This section details the 28 security practices identified through mining 861 security points (RQ1). We also present the perspective of the survey respondents about the usefulness of these security practices (RQ2). As described in Section \ref{subsec:identificationofsecuritypractices}, these 28 practices are classified into 6 groups. Similar to the arguments by Malavolta et al. \cite{malavolta2021mining}, the main goal behind revealing the usefulness of these security practices is to indicate their applicability in practice and assess the reliability of our analysis of the practices. We do not aim to rank these practices. Still, practitioners need to consider the design context and the requirements and constraints of their microservices systems and organizations when using these practices (see Section \ref{sec:recompractitioners}).

{\renewcommand{\arraystretch}{1.5}
\begin{table*}[]
\caption{Security practices for \textbf{authorizing and authentication} and the survey responses (in $\%$). \textbf{AU}: Absolutely Useful, \textbf{U}: Useful, \textbf{NU}: Not Useful, \textbf{ANU}: Absolutely Useful, \textbf{IDK}: I Don't Know, \textbf{MED}: Median, \textbf{AVG}: Average}
\label{group1}
\resizebox{\textwidth}{!}{
\begin{tabular}{l p{10cm} | c | c c c c c | c c } 
\hline
              &  \textbf{Security practices for authorizing and authentication}   &     \textbf{Sources}         & \textbf{AU} & \textbf{U} & \textbf{NU} & \textbf{ANU} & \textbf{IDK} & \textbf{MED} & \textbf{AVG} \\ \Xhline{3\arrayrulewidth}
\textbf{PA1} & Add an \say{identity microservice}, and authorize microservices access through \say{identity microservice} and token of each microservice. Each microservice has its own token, which is passed in each request between the microservices.              &

\begin{tabular}[c]{@{}c@{}} 
\cite{practicePA1andPA2andPT2}, \cite{practicePA1_2}, \cite{practicePA1_3}, \cite{practicePA1_4},\\ \cite{practicePA1_5}, \cite{practicePA1_6},  \cite{practicePA1_7}, \cite{practicePA1_8} \end{tabular}
&

\cellcolor[HTML]{C9FFAB} 32.43        &
\cellcolor[HTML]{B5FD8E} 45.95        & 
\cellcolor[HTML]{E0FFD1} 16.22        & 
\cellcolor[HTML]{FBFFF9} 1.35         & 
\cellcolor[HTML]{F9FFF7} 4.05         & 3            & 3.14         \\
\textbf{PA2} & Each microservice in the microservices architecture must be responsible for its own security, i.e., each microservice must have security enabled. &           
\cite{practicePA1andPA2andPT2}, \cite{practicePA2_2}, \cite{practicePA2_3}, \cite{practicePA4_2}  
&    
\cellcolor[HTML]{B5FD9C} 40.54        & 
\cellcolor[HTML]{C9FFAB} 33.78        & 
\cellcolor[HTML]{DBFFCE} 17.57        & 
\cellcolor[HTML]{EDFDE8} 6.76         & 
\cellcolor[HTML]{FBFFF9} 1.36        & 3            & 3.10         \\
\textbf{PA3} & Suppose you use an API Gateway approach in the microservices architecture. In that case, you do not need to have each microservice security enabled (you do not need to implement \textbf{PA2}) because the internal microservices can be protected by not being published out of the Docker Host. &               \begin{tabular}[c]{@{}c@{}} \cite{practicePA3andPA4}  \end{tabular}
& 
\cellcolor[HTML]{CBFFBA} 28.39        & 
\cellcolor[HTML]{CEFDBF} 24.32        & 
\cellcolor[HTML]{CFFCC1} 24.32        & 
\cellcolor[HTML]{E5FFDD} 12.16        & 
\cellcolor[HTML]{E9FFE2} 10.81        & 3            & 2.77         \\
\textbf{PA4} & A large microservices system is recommended to use an API Gateway approach for securing$/$authorizing$/$routing microservices.                                                                                                                                             &    
\cite{practicePA3andPA4}, \cite{practicePA4_2}, \cite{practicePA4_3andPA7}, \cite{practicePA4_5}, \cite{practicePA4_4}         
& 
\cellcolor[HTML]{8FFF52} 70.27        & 
\cellcolor[HTML]{CBFFBA} 25.68        & 
\cellcolor[HTML]{F5FFF2} 4.05         & 
\cellcolor[HTML]{FFFFFF} 0.00         & 
\cellcolor[HTML]{FFFFFF} 0.00         & 4            & 3.66         \\
\textbf{PA5} & Set an authorization boundary on each microservice even when (1) microservices are “internal” and (2) it is possible to set the authorization at the API Gateway level.                                                                                                                   &  
\cite{practicePA5andPT5_3}   
&
\cellcolor[HTML]{A2FF86} 45.95       & 
\cellcolor[HTML]{AEFF96} 40.54       & 
\cellcolor[HTML]{EBFFE4} 12.16        & 
\cellcolor[HTML]{FFFFFF} 0.00      & 
\cellcolor[HTML]{FBFFF9} 1.35         & 3            & 3.34      
\\
\textbf{PA6} & Use Public Key Infrastructure (PKI) signing$/$verification system to prevent round trips to the Authorization Service.                                                                                                          &   
\cite{practicePA8}          
& 
\cellcolor[HTML]{CAFFB6} 21.63       & 
\cellcolor[HTML]{A8FF89} 47.30       & 
\cellcolor[HTML]{E0FFD1} 14.86       & 
\cellcolor[HTML]{FBFFF9} 1.35        & 
\cellcolor[HTML]{E5FFDB} 14.86       & 3            & 3.05            \\ \hline
\end{tabular}}
\end{table*}}

\subsubsection{Authorization and Authentication}\label{Authorization-and-Authentication}

This group includes 6 practices for authorizing microservices or authenticating users in microservices systems (see Table \ref{group1}).

\faLightbulbO{} \textsc{\textbf{PA1.}} \textit{Add an “identity microservice”, and authorize microservices access through “identity microservice” and token of each microservice. Each microservice has its own token, which is passed in each request between the microservices.}
A microservices system includes several microservices, and each of the microservices performs specific tasks. Mostly, they need to communicate with each other to reach out to a target. Hence, authorizing microservices is an important task in these systems. Assume a user wants to be authenticated once and then access all the relevant microservices and client apps with their protected resources. This process is referred to as Single Sign-On (SSO) \cite{pereira2021security}. \textsc{\textbf{PA1}} is a way to implement SSO \cite{practicePA1andPA2andPT2}. In this practice, a microservice called “identity microservice” will check the token of microservices related to the user and make sure they are valid microservices. In case they are not valid microservices, the user will be redirected for the authentication.

This practice may negatively affect \textit{performance} due to too many round trips to the authentication server. Also, it affects \textit{scalability} because of the increased number of microservices. 58 (78.38\%) out of the 74 survey respondents confirmed that this practice is \textit{(absolutely) useful}. Some of the comments that the participants posted to support or refuse this practice are: 

\begin{myquote}
    \faThumbsOUp  ~ \say{\textit{[It is] useful to \textbf{maintain traceability} in requests between microservices.}} (Architect)
    
    \faThumbsOUp  ~ \say{\textit{I believe that microservices should have one service that is dedicated to authentication. Authentication service may be developed inside or subscribed as a service from other companies. Default gateway routes all incoming requests to the authentication service then the \textbf{authentication service validates whether the request has a valid token or not.} In my experience, each microservice has a private key of JWT, which checks all incoming requests before allowing access to a resource.}} (Software Engineer)

    \faThumbsODown  ~ \say{\textit{[I am] not sure [about] the \textbf{latency impact} [of this practice].}} (DevOps Engineer)

    \faThumbsODown  ~ \say{\textit{When working with microservices, it is important to remember that you have to follow your good practices, if you pass the token to all microservices, they will have to call the microservice for authorization always, in this case it is \textbf{faster and safer to use middleware} in the API gateway, this way centralizes authentication.}} (Software Engineer)
    
\end{myquote}

\faLightbulbO{} \textsc{\textbf{PA2.}} \textit{Each microservice in the microservices architecture must be responsible for its own security, i.e., each microservice must have security enabled.} Another way to implement SSO is to use a decentralized authentication protocol like OpenID, with which each microservice can handle its own security \cite{practicePA1andPA2andPT2}. In other words, with this practice, the user will have to authorize each microservice individually. More than half of the respondents (74.32\%) acknowledged that the practice is \textit{absolutely useful} or \textit{useful}. The following are 4 examples of the participants' comments that support \textsc{\textbf{PA2}}.

\begin{myquote}
    \faThumbsOUp ~ \say{\textit{Each microservice should enable security and have to \textbf{check all incoming requests before allowing access to the resource}, but [other stuff related to] microservice security such as token creation, refresh token, OAuth implementation should be handled by a separately dedicated authentication service.}} (Software Engineer)
    
    \faThumbsOUp ~ \say{\textit{Each service needs to be \textbf{protected with security token}.}} (Developer)
    
    \faThumbsOUp ~ \say{\textit{Absolutely useful \textbf{as long as microservices don't bother each other with different security protocols.} In my company we used to secure every module separately to ensure powerful security while maintaining their best performance.}} (Developer)
 
    \faThumbsOUp ~ \say{\textit{Enabling security between microservices - \textbf{this depends on the use case}. But \textbf{it is additional overhead}. If we are isolated from APIGW and the microservices needs to communicate with any microservices and if protected, then only it is advisable to have the security between microservices.}} (Architect)
\end{myquote}

On the other hand, some respondents mentioned that it is not necessary to secure all microservices: 

\begin{myquote}
    \faThumbsODown ~ \say{\textit{\textbf{[It is] not necessarily all microservices} in the architecture should be concerned with security, as this should be in the gateway API or in the infrastructure layer.}} (Software Engineer)

\end{myquote}

\faLightbulbO{} \textsc{\textbf{PA3.}} \textit{Suppose you use an API Gateway approach in the microservices architecture. In that case, you do not need to have each microservice security enabled (you do not need to implement \textsc{\textbf{PA2}}) because the internal microservices can be protected by not being published out of the Docker host.} One of the key challenges for practitioners is to decide to authorize microservices at the microservices level only, authorize at the API Gateway level, or both of them \cite{practicePA3andPA4}. Suppose they choose the API Gateway approach for this purpose. In that case, the microservices could only be accessed by other Containers within the Docker host through the \say{internal port} of each Container. 39 survey participants (52.71\%) considered this practice \textit{(absolutely) useful}. On the other hand, 27 (36.48\%) opposed it (24.32\% rated it as \textit{not useful}, and 12.16\% rated it as \textit{absolutely not useful}). The following is a positive comment that we received from the respondents.

\begin{myquote}
\faThumbsOUp ~ \say{\textit{
If we make the \textbf{security check at api gateway level}, then the microservices behind the gateway can \textbf{ignore token validity check} request to auth server. This will \textbf{reduce overhead}. Though they need to \textbf{check the access level for maintaining Role-Based Access Control (RBAC)}.}} (Technical Lead)

\end{myquote}

Some participants indicated that authorizing microservices at the microservices level and the API Gateway level is required for microservices systems as there should be zero trust security in such systems.

\begin{myquote}
    \faThumbsODown ~ \say{\textit{You can't know who is making the requests and you should have authentication/authorization enabled (\textbf{zero trust})}} (Software Engineer)
    
    \faThumbsODown ~ \say{\textit{Having an API gateway does not mean to disable security in microservices. In fact API gateway is used for \textbf{coarse grained security issues}, microservices are used for \textbf{fine grained security}. Both levels of security are required.}} (Requirements Engineer)
    
    \faThumbsODown ~ \say{\textit{It is important to \textbf{secure communication between microservices} even if you use API Gateway.}} (Technical Lead)
    
\end{myquote}

\faLightbulbO{} \textsc{\textbf{PA4.}} \textit{A large microservices system is recommended to use an API Gateway approach for securing/authorizing/routing microservices.} 
The API Gateway approach causes overhead in small microservices systems and needs too many steps when coding and updating features \cite{practicePA4_2, practicePA3andPA4}. Out of 28 practices, \textbf{PA4} was mostly rated by the survey participants as \textit{absolutely useful} or \textit{useful} (more than 95\%). No one voted this practice as \textit{absolutely not useful}. 

We received only positive comments on this practice, indicating that it works well for routing, minimizes coupling, and supports the evolution of microservices.

\begin{myquote}
    
    \faThumbsOUp ~ \say{\textit{[This practice] \textbf{simplify code and is robust}}}
    
    \faThumbsOUp ~ \say{\textit{I agree default \textbf{gateway [can be used] for routing} but authentication and authorization should be handled by separate service.}} (Software Engineer)
    
    \faThumbsOUp ~ \say{\textit{Using gateway makes it easier for the client because (1) they don’t have to deal with many different addresses of each service, (2) abstraction is highly enforced which also enhances security because \textbf{endpoints of the services are not directly exposed}, thereby making it difficult for attackers, and (3) \textbf{coupling is minimised}.}} (Developer)

    \faThumbsOUp ~ \say{\textit{We are using it, and \textbf{it enables routing correctly}.}} (Technical Lead)

\end{myquote}

\faLightbulbO{} \textsc{\textbf{PA5.}} \textit{Set an authorization boundary on each microservice even when (1) microservices are “internal” and (2) it is possible to set the authorization at the API Gateway.} This practice is recommended when developers want to add authorization with Ocelot\footnote{\url{https://ocelot.readthedocs.io/en/latest/index.html\#}} \cite{practicePA5andPT5_3}. However, developers need to balance between security and simplicity. The majority of our survey respondents agreed with the usefulness of this practice (86.49\% rated it as \textit{absolutely useful} or \textit{useful}).

As shown in the following comments, \textbf{PA5} (1) provides a native cloud solution, (2) is useful for accidental wrong access/modification, and (3) reduces the security concerns in internal microservices.

\begin{myquote}
    \faThumbsOUp ~ \say{\textit{It uses a \textbf{native cloud solution}.}} (DevOps Engineer)
    
    \faThumbsOUp ~ \say{\textit{Authorisation boundary is very important and useful not only in terms of security but also for \textbf{accidental wrong access/modification}.}} (Software Engineer)
    
    \faThumbsOUp ~ \say{\textit{[It] is cool because you create an \textbf{outer layer of security} and your internal services don't have to worry about security.}} (Software Engineer)
    
\end{myquote}   

In contrast, a detractor mentioned that:

\begin{myquote}
    \faThumbsODown ~ \say{\textit{Some requests navigate through many services before returning a response, so implementing authorization in API Gateway, I believe, is not the best practice. The default gateway should only do the \textbf{routing stuff}, adding some security-related task to the default gateway is not a \textbf{scalable solution}. Because if you want to implement SSO in the future, I think it's a bit \textbf{difficult to implement} it.}} (Software Engineer)
\end{myquote}

\faLightbulbO{} \textsc{\textbf{PA6.}} \textit{Use Public Key Infrastructure (PKI) signing/verification system to prevent round trips to the Authorization Service.} This practice aims to handle the authorization in a microservices environment \cite{practicePA8}. Most respondents (68.93\% considered this practice \textit{useful} or \textit{absolutely useful}. 
As an example of the positive comments from the respondents on this practice, we have:

\begin{myquote}

    \faThumbsOUp ~ \say{\textit{If we use PKI then \textbf{each microservice can validate the security tokens} instead of sending a request to identity microservice to validate the tokens.}} (Developer)
    
\end{myquote}

Detractors noted that the use of PKI in a microservices architecture is costly.

\begin{myquote}

    \faThumbsODown ~ \say{\textit{Not useful, depending on the number of requests between microservices, performing key verification \textbf{can be very costly}.}} (Architect)
    
    \faThumbsODown ~ \say{\textit{This will have a \textbf{high cost of development and maintenance}.}} (DevOps Engineer)
    
\end{myquote}

{\renewcommand{\arraystretch}{1.5}
\begin{table*}[]
\centering
\caption{Security practices for \textbf{tokens and credentials} and the survey responses (in \%). \textbf{AU}: Absolutely Useful, \textbf{U}: Useful, \textbf{NU}: Not Useful, \textbf{ANU}: Absolutely Useful, \textbf{IDK}: I Don't  Know, \textbf{MED}: Median, \textbf{AVG}: Average}
\label{group2}
\resizebox{\textwidth}{!}{
\begin{tabular}{l p{9.5cm} | c | c c c c c | c c }

\hline
             &   \textbf{Security practices for tokens and credentials}       &         \textbf{Sources}                                                            & \textbf{AU} & \textbf{U} & \textbf{NU} & \textbf{ANU} & \textbf{IDK} & \textbf{MED} & \textbf{AVG} \\ \Xhline{2.3\arrayrulewidth}
\textbf{PT1} & Use a method based on the Public/Private key to secure microservices through JSON Web Token (JWT).                          &  \begin{tabular}[c]{@{}c@{}}\cite{practicePT1}, \cite{practicePT1_2}, \cite{practicePT1_3}, \\ \cite{practicePT1_4}, \cite{practicePT1_5}\end{tabular}  
&
\cellcolor[HTML]{B5FF99} 39.19        & 
\cellcolor[HTML]{A1FF7E} 51.35       & 
 \cellcolor[HTML]{F9FFF7} 2.70        & 
 \cellcolor[HTML]{FFFFFF} 0.00            & 
 \cellcolor[HTML]{F3FFF1} 6.76          & 3               & 3.39                  \\     
\textbf{PT2} & In microservices systems, use JSON Web Tokens (JWTs) to handle the session expiration/revocation.                  &      
 \cite{practicePA1andPA2andPT2},  \cite{practicePT2_2}   
&
\cellcolor[HTML]{B5FD9C} 40.54        &
\cellcolor[HTML]{A3FF82} 48.65       & 
\cellcolor[HTML]{F5FEF2} 4.05         & 
\cellcolor[HTML]{FFFFFF} 0.00            & 
\cellcolor[HTML]{EDFDE8} 6.76          & 3               & 3.39                   \\   
\textbf{PT3} & Secure caches of credentials in each microservice that needs to access other microservices.                       &   \cite{practicePT3}       & 
\cellcolor[HTML]{E5FFDD} 14.87        & 
\cellcolor[HTML]{A8FF89} 43.24         & 
\cellcolor[HTML]{D4FFC7} 20.27        & 
\cellcolor[HTML]{EBFFE4} 10.81          & 
\cellcolor[HTML]{EBFFE4} 10.81          & 3               & 2.70                \\ 
\textbf{PT4} & Decode JSON Web Token (JWT) at the microservices level instead of the API Gateway level.                              &          \cite{practicePT4}     & 
\cellcolor[HTML]{C3FEB1} 27.03        & 
\cellcolor[HTML]{BDFFA6} 33.78       & 
\cellcolor[HTML]{E0FFD1} 18.92        & 
\cellcolor[HTML]{EDFDE8} 6.76          & 
\cellcolor[HTML]{E4FFDC} 13.51         & 3              & 2.94                       \\ 
\textbf{PT5} & Endpoints of microservices like server information, health check, and logging level must be secured in the request/response chain. &     
\cite{practicePT5}, \cite{practicePT5_2}, \cite{practicePA5andPT5_3}        
&
\cellcolor[HTML]{B6FFA0} 36.49        & 
\cellcolor[HTML]{BDFFA6} 35.14       & 
\cellcolor[HTML]{DBFFCE} 18.91          & 
\cellcolor[HTML]{F9FFF7} 2.70          & 
\cellcolor[HTML]{F3FFF1} 6.76          & 
3               &
3.13                   \\ \hline
\end{tabular}}
\end{table*}}

\subsubsection{Token and Credentials}\label{Token-and-Credentials}

As shown in Table \ref{group2}, this group includes 5 practices for handling sensitive information in a microservices system.

\faLightbulbO{} \textsc{\textbf{PT1.}} \textit{Use a method based on the Public/Private key to secure microservices through JSON Web Token (JWT).} 
JWTs can be signed in a microservices system and generate 2 key pairs (private signing key and public verification key) \cite{jwt}. The public verification key generated by a JWT can be distributed to all microservices in the microservices system. If microservice A wants to decrypt the information in microservice B, it only needs to know the private signing key created by microservice B \cite{practicePT1}. If the microservices system uses the API Gateway approach, the API Gateway should also know the private signing key.

90.54\% (67) of the survey respondents rated it as \textit{absolutely useful} or \textit{useful}. Only 2 practitioners chose \textit{not useful}. This practice received only positive comments, in which the respondents offered to use the OAuth stream in addition to JWT. 

\begin{myquote}
    \faThumbsOUp ~ \say{\textit{JWT alone is not enough. [It is] interesting to use an \textbf{OAuth stream} with client credentials + cookie, in a Gateway API strategy.}} (Architect)
    
    \faThumbsOUp ~ \say{\textit{For external facing APIs, we can use \textbf{OAuth based authentication}.}} (DevOps Engineer)
\end{myquote}

\faLightbulbO{} \textsc{\textbf{PT2.}} \textit{In microservices systems, use JSON Web Tokens (JWTs) to handle the session expiration/revocation.} This practice recommends using Redis tool\footnote{\url{https://redis.io}} to track token revocations \cite{practicePA1andPA2andPT2}. \textbf{PT2} received almost similar positive feedback to \textbf{PT1} from the survey respondents (89.19\% \textit{absolutely useful} or \textit{useful}). A respondent stated that this practice is useful if someone uses JWT for communication between microservices behind the gateway. Another participant confirmed \textbf{PT2} as a useful practice, but he/she mentioned that JWT is not the only method to handle the session expiration and revocation.

\begin{myquote}
    \faThumbsOUp ~ \say{\textit{Communication between \textbf{microservices behind the gateway} can be \textbf{JWT} which is a value token. But from \textbf{client to gateway} should be a \textbf{reference token} which does not contain any sensitive information.}} (Architect)
    
    \faThumbsOUp ~ \say{\textit{JWT is \textbf{not} the only \textbf{option}.}} (Software Engineer)
\end{myquote}

\faLightbulbO{} \textsc{\textbf{PT3.}} \textit{Secure caches of credentials in each microservice that needs to access other microservices.} 
More than 55\% of the practitioners verified that \textbf{PT3} \cite{practicePT3} is \textit{absolutely useful} or \textit{useful}, while 31.08\% rated it as \textit{not useful} or \textit{absolutely not useful}. A software engineer with more than 3 years of experience in security of microservices systems pointed out:

\begin{myquote}

    \faThumbsOUp ~ \say{\textit{If the cached credentials \textbf{are not secured}, the entire credential system is \textbf{questionable}.}} (Software Engineer)
    
\end{myquote}

Some respondents believed that (1) the usefulness of \textbf{PT3} depends on whether the microservice is stateful or stateless, and (2) the overhead of securing credentials.

\begin{myquote}

    \faThumbsODown ~ \say{\textit{It depends on the [micro]service type weather it is \textbf{stateful} or \textbf{stateless}. If the service is stateful, we may think about a way how to store the credential and use it in the next request. Else we use a private key to validate the request token.}} (Software Engineer)

    
    \faThumbsODown ~ \say{\textit{\textbf{Security is necessary but an overhead}, only secure what needs securing. Maintain separation of concerns, \textbf{manage/cache user credentials in one dedicated service}.}} (Technical Lead)

\end{myquote}

\faLightbulbO{} \textsc{\textbf{PT4.}} \textit{Decode JSON Web Token (JWT) at the microservices level instead of the API Gateway level.} 
45 participants (60.81\%) believed that JWTs should be decoded at the microservices level because they mostly include relevant information for authentication and authorization. API Gateway can manage the JWTs in the form of Fail-fast (a.k.a. fail early), and it is just recommended to verify access tokens at the microservices level \cite{practicePT4}.  Our analysis shows that 19 respondents (25.68\%) did not agree with the usefulness of \textbf{PT4}. Below are 3 comments that question \textbf{PT4} and rationalize why decoding JWTs should be done at the API Gateway level.


\begin{myquote}
    \faThumbsODown ~ \say{\textit{Since JSON token is decoded once at the gateway level, \textbf{the overhead of each service having to deal with the decoding is removed}, thereby making availability better.}} (Developer)
\end{myquote}


\begin{myquote}
    \faThumbsODown ~ \say{\textit{I believe decoding can be done \textbf{at the API Gateway level} and from there, the request should opt for a different way to hit the internal services. That will free the services from doing any kind of decoding work. I think it would have \textbf{better performance and be more secure}.}} (Architect)
    
    \faThumbsODown ~ \say{\textit{I think that JWT tokens can be checked at the API Gateway, \textbf{but not 'instead of' the microservice level}. Checking at the gateway will keep failed authentications away from any unnecessary processing but they should be verified by the Microservice.}} (Technical Lead)
\end{myquote}

\faLightbulbO{} \textsc{\textbf{PT5.}} \textit{Endpoints of microservices like server information, health check, and logging level must be secured in the request/response chain.}
Our analysis shows that server information, health check, and logging level contain sensitive information and should be secured as part of securing a microservices system \cite{practicePT5, practicePT5_2}. Depending on the requirements of a microservices system, developers may only use some of these endpoints. 71.63\% of the respondents opted for \textit{absolutely useful} or \textit{useful} for this practice. Some comments that indicate the importance of securing endpoints are shown as:

\begin{myquote}
     \faThumbsOUp ~ \say{\textit{Leaving \textbf{diagnostics information} unsecured may expose loopholes in the system, which makes it easier for attackers. Again sensitive information can be leaked to unauthorized users.}} (Developer)
     
      
    \faThumbsOUp ~ \say{\textit{All endpoints, including \textbf{diagnostic endpoints, must be secured}. These are prone to attacks and can leak potentially sensitive data.}} (Technical Lead)
      
    \faThumbsOUp ~ \say{\textit{Specially \textbf{logs could contain data} that need to be protected at all time.}} (DevOps Engineer)
\end{myquote}

In contrast, a respondent disagreed with protecting health checks as it may not allow the implementation of a fault tolerance strategy.

\begin{myquote}
    \faThumbsODown ~ \say{\textit{\textbf{Health checks} should \textbf{not} be protected, in a fault tolerance strategy whoever makes the request needs to know if the microservice is active, to allow a retry alternative.}} (Architect)
\end{myquote}

{\renewcommand{\arraystretch}{1.5}
\begin{table*}[]
\caption{Security practices for \textbf{internal and external microservices} and the survey responses (in \%). \textbf{AU}: Absolutely Useful, \textbf{U}: Useful, \textbf{NU}: Not Useful, \textbf{ANU}: Absolutely Useful, \textbf{IDK}: I Don't  Know, \textbf{MED}: Median, \textbf{AVG}: Average}
\label{group3}
\resizebox{\textwidth}{!}{
\begin{tabular}{l p{9.5cm} | c | c c c c c | c c }
\hline
             &       \textbf{Security practices for internal and external microservices}      &   \textbf{Sources}                                                                                                                                                   & \textbf{AU} & \textbf{U} & \textbf{NU} & \textbf{ANU} & \textbf{IDK} & \textbf{MED} & \textbf{AVG}  \\ \Xhline{2.3\arrayrulewidth}
\textbf{PI1} & Developers can use internal microservices secured with a different token than external microservices. In this scenario, API Gateway acts as a token issuer for the internal microservices.                                          &              \cite{practicePI1andPI2}     & 
\cellcolor[HTML]{CEFDBF} 27.03        & 
\cellcolor[HTML]{AAFF90} 43.24       & 
\cellcolor[HTML]{E0FFD1} 16.22        & 
\cellcolor[HTML]{FBFFF9} 2.7          & 
\cellcolor[HTML]{E9FFE2} 10.81         & 3               & 3.06                 \\ 
\textbf{PI2} & Developers can use internal microservices secured using the tokens of external microservices, and their permission must be controlled using Access Control List (ACL). In this scenario, API Gateway forwards the tokens to the internal microservices. &  \cite{practicePI1andPI2}
&
\cellcolor[HTML]{E0FFD1} 17.57        & 
\cellcolor[HTML]{ACFF93} 43.24       & 
\cellcolor[HTML]{DFFFD5} 13.51        & 
\cellcolor[HTML]{F5FEF2} 6.76          & 
\cellcolor[HTML]{DBFFCE} 18.92           & 3               & 2.88                       \\
\textbf{PI3} & Whether microservices are only internally used within an organization or are externally accessible to third parties, authentication is required either way.                                                                                            &   \cite{practicePI3}, \cite{practicePI3_2}, \cite{practicePI3_3}
&
\cellcolor[HTML]{9AFF7B} 54.05        & 
\cellcolor[HTML]{C3FEB1} 29.73       & 
\cellcolor[HTML]{EEFFE9} 6.76         & 
\cellcolor[HTML]{F5FEF2} 5.41          & 
\cellcolor[HTML]{F5FEF1} 4.05          & 4               & 3.38                     \\
\textbf{PI4} & In an internal microservice use case, “client credential” should not get exposed to the third party.                                                                                                                                &   \cite{practicePI4}                & 
\cellcolor[HTML]{95FE76} 56.76          & 
\cellcolor[HTML]{B6FFA0} 35.14       & 
\cellcolor[HTML]{F5FEF2} 4.05         & 
\cellcolor[HTML]{FBFFF9} 1.35          & 
\cellcolor[HTML]{F9FFF7} 2.70          & 4               & 3.51                      \\ 
\textbf{PI5} & Microservices systems made of components should be isolated and internal calls should not be leaked outside their boundaries.                                                                                                          &    \cite{practicePP1_PP2_PI5}                 & 
\cellcolor[HTML]{9BFF7E} 51.35        & 
\cellcolor[HTML]{B0FF99} 40.54       & 
\cellcolor[HTML]{F3FFF1} 5.41         & 
\cellcolor[HTML]{FFFFFF} 0.00            & 
\cellcolor[HTML]{F9FFF7} 2.70          & 4               & 3.47                      \\ 
\textbf{PI6} & Encrypt tokens if they are going to be exposed to the outside of the system boundary.                                                                                                                                         &   \cite{practicePI6}                      & 
\cellcolor[HTML]{95FE76} 55.41          & 
\cellcolor[HTML]{E5FFDE} 22.97       & 
\cellcolor[HTML]{E5FFDD} 12.16        & 
\cellcolor[HTML]{F5FEF2} 4.05          & 
\cellcolor[HTML]{F5FEF2} 5.41          & 4               & 3.37                    \\ 
\textbf{PI7} & It is recommended to minimize the number of HTTP dependencies between internal microservices. This will minimize the future impact on microservices performance and Denial-of-Service attacks.                           &    \cite{practicePI7andPD1}                          & 
\cellcolor[HTML]{B3FF9F} 37.84        & 
\cellcolor[HTML]{B0FF99} 39.19       & 
\cellcolor[HTML]{E4FFDC} 13.51        & 
\cellcolor[HTML]{FFFFFF} 0.00            & 
\cellcolor[HTML]{EBFFE4} 9.46          & 3               & 3.27                       \\ \hline
\end{tabular}}
\end{table*}}

\subsubsection{Internal and External Microservices}\label{Internal-and-External-Microservices}
This group of practices focuses on securing a set of microservices. Part of these microservices (internal microservices) is used inside an organization, and the rest (external microservices) may be used by any third-party (see Table \ref{group3}).

\faLightbulbO{} \textsc{\textbf{PI1.}} \textit{Developers can use internal microservices secured with a different token than external microservices. In this scenario, API Gateway acts as a token issuer for the internal microservices.} 52 respondents (70.27\%) acknowledged that developers could use unique tokens for securing internal and external microservices. Two practitioners rated it \textit{absolutely not useful} (2.7\%), and 12 practitioners considered it \textit{not useful} (16.22\%). Below are 2 negative comments on this practice. 

\begin{myquote}
    \faThumbsODown ~ \say{\textit{The default gateway should not handle \textbf{Identity and Access Management (IAM)} task.}} (Software Engineer)

    \faThumbsODown ~ \say{\textit{There is \textbf{no need to use token and secure} internal microservices as long as they are not accessible from outside.}} (Software Engineer)
\end{myquote}

\faLightbulbO{} \textsc{\textbf{PI2.}} \textit{Developers can use internal microservices secured using the tokens of external microservices, and their permission must be controlled using Access Control List (ACL). In this scenario, API Gateway forwards the tokens to the internal microservices.} Similar to \textbf{PI1}, \textbf{PI2} aims to secure internal microservices. However, it uses the tokens of external microservices for securing internal microservices \cite{practicePI1andPI2}. 60.81\% of the participants stated that \textbf{PI2} is \textit{absolutely useful} or \textit{useful}. 20.27\% rated this practice as \textit{not useful} or \textit{absolutely not useful}. The following comment includes negative feedback on this practice.

 \begin{myquote}
    \faThumbsODown ~ \say{\textit{I don't know how useful is to having the same token internally and externally and \textbf{encrypting the token is always the best}.}} (Architect)
    
    \faThumbsODown ~ \say{\textit{External authorization and Internal authorization are different. \textbf{External token must be used to authorize the user}. \textbf{Don't mix}.}} (Architect)
\end{myquote}

\faLightbulbO{} \textsc{\textbf{PI3.}} \textit{Whether microservices are only internally used within an organization or are externally accessible to third parties, authentication is required either way.} The majority of the survey respondents (83.78\%, 62 out of 74) considered \textbf{PI3} as \textit{absolutely useful} or \textit{useful}. They argued that the authenticated microservices remain secure in the following scenarios: (1) the occurrence of misconfigurations that lead to exposure of internal microservices to outside, and (2) if a security hole is opened in the firewall \cite{practicePI3}.

\begin{myquote}

    \faThumbsOUp ~ \say{\textit{If authentication is not enabled for internal microservices, then as soon as the \textbf{internal physical network gets compromised}, the entire microservices system is compromised, which is a disaster.}} (Software Engineer)
    
\end{myquote}

On the other hand, some respondents believed that this practice would be only necessary or useful under certain circumstances.

\begin{myquote}    
    \faThumbsODown ~ \say{\textit{It \textbf{depends on the service} that the microservice gives. Some services may need authentication, and some may not need a user to authenticate.}} (Developer)
    
    \faThumbsODown ~ \say{\textit{By having communication via the [message] broker, it is not necessary to authenticate in internal microservices, but \textbf{if you use REST in microservices}, this is making a bad practice, and in that case, \textbf{it will be necessary}.}} (DevOps Engineer)
\end{myquote}

\faLightbulbO{} \textsc{\textbf{PI4.}} \textit{In an internal microservice use case, “client credential” should not get exposed to the third party.} The client credentials are identifiers for accessing client data. It is strongly advised to distinguish internal client credentials from external ones \cite{practicePI4}. This practice was rated as \textit{absolutely useful} or \textit{useful} by more than 90\% participants. Only one respondent noted that if the third party is valid for the internal microservices, there is no problem exposing client credentials to the third party.

\begin{myquote}
    \faThumbsODown ~ \say{\textit{If we are talking about the grant type client credentials in OAuth2 and the \textbf{client ID/secret identifies the third party} system, I don't see a problem exchanging the “client credential” with the third party.}} (Requirements Engineer)
    
\end{myquote}

\faLightbulbO{} \textsc{\textbf{PI5.}} \textit{Microservices systems made of components should be isolated and internal calls should not be leaked outside their boundaries.} This practice has more focus on controlling the components' exposure. Most participants (91.89\%, 68 out of 74) marked it as \textit{absolutely useful} or \textit{useful}. Similar to \textbf{PI4}, none of the participants rated this practice as \textit{absolutely not useful}.

\faLightbulbO{} \textsc{\textbf{PI6.}} \textit{Encrypt tokens if they are going to be exposed to the outside of the system boundary.} From the perspective of a software developer with 10 years of experience, it is needed to encrypt the tokens because they contain some authorization-related information \cite{practicePI6}. This practice was \textit{absolutely useful} for 41 respondents (55.41\%) and \textit{useful} for 17 respondents (22.97\%). The following are 2 positive comments which state that the tokens should be encrypted at all times.

\begin{myquote}
    \faThumbsOUp ~ \say{\textit{Tokens can \textbf{be hacked} so should \textbf{always} be encrypted.}} (Architect)

    \faThumbsOUp ~ \say{\textit{Tokens should \textbf{be encrypted with a public key} and get decrypted with a private key regardless of the fact that the internal service is receiving it or a client.}} (Architect)
\end{myquote}    

However, a survey respondent questioned the need for adding another layer of encryption if tokens are already signed.

\begin{myquote}
    \faThumbsODown ~ \say{\textit{Assuming tokens are \textbf{already signed}, what is the \textbf{reason} to have \textbf{another layer} of encryption?}} (Software Engineer)
\end{myquote}

\faLightbulbO{} \textsc{\textbf{PI7.}} \textit{It is recommended to minimize the number of HTTP dependencies between internal microservices. This will minimize the future impact on microservices performance and Denial-of-Service attacks.} 
Our analysis of the security points revealed that some developers advised that fewer communications between internal microservices are better because being autonomous and available to the client is one of the purposes of microservices. If we employ HTTP dependencies between microservices, it can violate the autonomy of microservices \cite{practicePI7andPD1}. It also impacts the performance of microservices when one of them does not perform well \cite{practicePI7andPD1}.

57 (77.03\%) participants considered \textbf{PI7} as \textit{absolutely useful} or \textit{useful}. The survey respondents pointed out that it would be useful to avoid HTTP calls as much as possible because they may create some problems for internal microservices calls. Furthermore, the respondents emphasized that HTTP dependencies should be reduced because it is against the separation of concerns principle in the design of microservices systems. They also recommended using gRPC instead of HTTP for synchronous calls. An alternative to prevent the Distributed Denial-of-Service (DDoS) attack is to use the Backends For Frontends (BFF) pattern\footnote{\url{https://samnewman.io/patterns/architectural/bff}}.

\begin{myquote}
    \faThumbsOUp ~ \say{\textit{HTTP calls are synchronous. Hence too much use of it for inter-service calls may \textbf{cause availability issues}. If indeed synchronous calls are required, then \textbf{gRPC} may be used}.} (Developer)
    
    \faThumbsOUp ~ \say{\textit{One way to prevent \textbf{DDoS} is to work with \textbf{Backends For Frontends (BFF)} and only expose it to the world, and not expose each of your microservices to be accessible by external HTTP requests.}} (Software Engineer)
    
    \faThumbsOUp ~ \say{\textit{\textbf{Separation of concerns} is a major feature of pure microservices.}} (Software Engineer)
    
    \faThumbsOUp ~ \say{\textit{This is required for \textbf{performance}.}}
    
\end{myquote}

A negative comment that we received on this practice is:

\begin{myquote}
    \faThumbsODown ~ \say{\textit{When we came to microservice architecture, \textbf{the most used way of synchronous messaging between services was by using HTTP} so as many requests could be sent and received to do the task. I don't recommend minimizing the number of requests as a solution.}} (Developer)
\end{myquote}

{\renewcommand{\arraystretch}{1.5}
\begin{table*}[]
\caption{Security practices for \textbf{microservices communications} and the survey responses (in \%). \textbf{AU}: Absolutely Useful, \textbf{U}: Useful, \textbf{NU}: Not Useful, \textbf{ANU}: Absolutely Useful, \textbf{IDK}: I Don't  Know, \textbf{MED}: Median, \textbf{AVG}: Average}
\label{group4}
\resizebox{\textwidth}{!}{
\begin{tabular}{l p{9.5cm} | c | c c c c c | c c }
\hline
             &    \textbf{Security practices for microservices communications}   & \textbf{Sources}                                                                                                                                                     & \textbf{AU} & \textbf{U} & \textbf{NU} & \textbf{ANU} & \textbf{IDK} & \textbf{MED} & \textbf{AVG}  \\ \Xhline{2.3\arrayrulewidth}
\textbf{PM1} & Use OAuth2 “Client Credentials Flow” if two microservices that trust each other want to talk together from the backends.                                         & \cite{practicePM1}, \cite{practicePA6}, \cite{practicePM1_2}, \cite{practicePM1_3}              &
\cellcolor[HTML]{DBFFCE} 24.32          & 
\cellcolor[HTML]{95FE76} 50.0         & 
\cellcolor[HTML]{F3FFF1} 6.76         & 
\cellcolor[HTML]{F9FFF7} 2.70          & 
\cellcolor[HTML]{E0FFD1} 16.22         & 3               & 3.15               \\ 
\textbf{PM2} & The connection between a microservice and its respective database should be protected by a security protocol, like Transport Layer Security (TLS).               &  \cite{practicePM2}, \cite{practicePM2_2}                  &
\cellcolor[HTML]{AAFF90} 48.65        & 
\cellcolor[HTML]{B3FF9F} 36.49       & 
\cellcolor[HTML]{EBFFE4} 8.11         & 
\cellcolor[HTML]{F9FFF7} 2.70          & 
\cellcolor[HTML]{F5FEF2} 4.05          & 4               & 3.37                    \\ 
\textbf{PM3} & When a microservice in a microservices architecture needs to call another microservice, the access token should be passed around microservices with the request. & \cite{practicePM3}
&
\cellcolor[HTML]{C9FFAB} 35.14        & 
\cellcolor[HTML]{B6FFA0} 37.84       & 
\cellcolor[HTML]{D5FFC9} 18.92        & 
\cellcolor[HTML]{F5FEF2} 4.05          & 
\cellcolor[HTML]{F5FEF2} 4.05          & 3               & 3.08              \\ 
\textbf{PM4} & It is recommended to use the gRPC framework for internal microservice-to-microservice synchronous communication.                                                 & \begin{tabular}[c]{@{}c@{}}\cite{practicePM1_3}, \cite{practicePM4_1}, \cite{practicePM4_2}, \cite{practicePM4_3},\\ \cite{practicePM4_4}, \cite{practicePD1_3}\end{tabular}
&
\cellcolor[HTML]{D5FFC9} 20.27        & 
\cellcolor[HTML]{B3FF9F} 37.84       & 
\cellcolor[HTML]{DBFFCE} 16.22          & 
\cellcolor[HTML]{FFFFFF} 0.00            & 
\cellcolor[HTML]{D4FFC7} 25.67         & 3               & 3.05              \\ \hline
\end{tabular}}
\end{table*}}

\subsubsection{Microservices Communications}\label{Microservices-Communications}
Table \ref{group4} represents 4 practices that are related to authenticating and authorizing requests when 2 or more microservices are communicating.

\faLightbulbO{} \textsc{\textbf{PM1.}} \textit{Use OAuth2 “Client Credentials Flow” if two microservices that trust each other want to talk together from the backends.} In such communications, there is no end-user identity involved. \textbf{PM1} emphasizes the trust between microservices where they explicitly call each other \cite{practicePM1}. Assume that there are a user and 2 microservices (A, B). The user accesses microservice A through a JSON Web Token (JWT). At this time, microservice A needs to access microservice B. The “OAuth2 client credentials grant” is recommended to handle the communication between these 2 microservices. If 2 microservices do not trust each other, the “OAuth2 client credentials grant” provides a good way to handle the authentication between these 2 microservices \cite{practicePA6}. In this case, each microservice will use its own credentials to obtain a token through the “token microservice” (i.e., a microservice that is responsible for generating, renewing, and validating a token) and use it to connect to another microservice.

A large number of the survey participants (i.e., more than 70\%) considered this as an (\textit{absolutely}) \textit{useful} practice. A few were the opposite of it (less than 10\%). As a positive comment on this practice, we have:

\begin{myquote}

    \faThumbsOUp  ~ \say{\textit{Client credential is only valid when the \textbf{intercommunication} does not specify the \textbf{current active user}.}} (Architect)
    
\end{myquote}

We only received a negative comment on this practice.

\begin{myquote}
    \faThumbsODown ~ \say{\textit{Intercommunications between microservices \textbf{should be a custom grant type}, not client credential.}} (Architect)
\end{myquote}

\faLightbulbO{} \textsc{\textbf{PM2.}} \textit{The connection between a microservice and its respective database should be protected by a security protocol, like Transport Layer Security (TLS).} Our analysis of the collected security points taught us that developers should be worried about the security of communication between a microservice and its database (or even other databases) \cite{practicePM2}. As an essential part of the data protection strategy, Microsoft strongly advises protecting data in transit \cite{encryptionpractices}. Moreover, because the data is exchanged from many locations, Secure Sockets Layer (SSL) or TLS protocols are highly recommended. 63 respondents admitted this practice (48.65\% \textit{absolutely useful} and 36.49\% \textit{useful}).

They mentioned that \textbf{PM2} is useful to prevent unauthorized access of microservices to the database. They also recommended that using the gRPC framework with the TLS protocol is a good practice when a microservice wants to communicate to its own database.

\begin{myquote}
    \faThumbsOUp  ~ \say{\textit{It is better to secure the connection of a service and its database because it \textbf{prevents unauthorized access} to the database.}} (Developer)
    
    \faThumbsOUp  ~ \say{\textit{It's good to use \textbf{gRPC framework and TLS} while accessing the DB.}} (Architect)
    
    \faThumbsOUp  ~ \say{\textit{TLS will ensure that \textbf{all traffic is encrypted}.}} (Technical Lead)
    
\end{myquote}

Some others argued that (1) there is no need to use \textbf{PM2} once the database is in a private network, and (2) adding a security protocol in a connection causes more complexity.

\begin{myquote}

    \faThumbsODown ~ \say{\textit{The DB must \textbf{be embedded} in the service container or in a private network. Therefore there is \textbf{no need to encrypt} a local connection.}} (Architect)
    
    \faThumbsODown  ~ \say{\textit{Adding SSL to database connection only \textbf{adds more complexity}.}} (Software Engineer)
    
\end{myquote}

\faLightbulbO{} \textsc{\textbf{PM3.}} \textit{When a microservice in a microservices architecture needs to call another microservice, the access token should be passed around microservices with the request.} Imagine there are a user and 2 microservices (A and B). The user and microservice A are in Scope A (i.e., the user is authorized for microservice A).  Microservice B is in Scope B (i.e., the user is not authorized for microservice B). Suppose the user with the access token wants to use a resource from microservice A and at the same time, microservice A must call microservice B to give the resource to the user. In that case, \textbf{PM3} is recommended to prevent any communication failure \cite{practicePM3}. 

72.98\% of the proponents agreed \textbf{PM3} is \textit{(absolutely) useful}. Two participants provided conditions for the usefulness of this practice in the following comments.

\begin{myquote}

\faThumbsOUp  ~ \say{\textit{Only if you are going to \textbf{get some information from the user}, otherwise, it [passing the access token around microservices with the request] is not necessary.}} (DevOps Engineer)

\faThumbsOUp  ~ \say{\textit{This is useful \textbf{when each microservice parses the token, gets the requester information and use that for some operation}. This strongly verifies that the owner of the session is doing that particular operation, rather that reading some parameters to identify who is the owner of the operation.}} (Technical Lead)

\end{myquote}

\faLightbulbO{} \textsc{\textbf{PM4.}} \textit{It is recommended to use the gRPC framework for internal microservice-to-microservice synchronous communication.} gRPC is a communication protocol using HTTP2 \cite{http2} and Protocol Buffers \cite{protocolbuffer}. The analysis of the security points indicated that gRPC provides an effective solution for direct synchronous communication between microservices \cite{practicePM4_1}, \cite{practicePM4_2}. 43 respondents agreed that \textbf{PM4} is a (\textit{absolutely}) \textit{useful} practice (58.11\%). More than 25\% of the participants were not familiar with this practice. No one selected \textit{absolutely not useful} for this practice.

Proponents argued that gRPC is faster than HTTP because it uses binary encoding. However, the costs of using it should be considered.

\begin{myquote}
    \faThumbsOUp  ~ \say{\textit{gRPC uses binary encoding, which makes it \textbf{faster}.}} (Developer)
    
    \faThumbsOUp  ~ \say{\textit{You can take \textbf{benefit over HTTP}. But it depends.}} (Technical Lead)
    
    \faThumbsOUp  ~ \say{\textit{gRPC is useful but it has a \textbf{cost}, you should think about its \textbf{advantages to implement} it.}} (Software Engineer)
    
\end{myquote}

In contrast, a few respondents pointed out that \textbf{PM4} is not useful and should be avoided because, e.g., it causes difficulties in development and debugging.

\begin{myquote}

    \faThumbsODown  ~ \say{\textit{This is an option, but should \textbf{not be a requirement}.}} (Software Engineer)

    \faThumbsODown  ~ \say{\textit{Binary data transfer makes development and debugging \textbf{very difficult} and does \textbf{not add much of value} in terms of security or network performance.}} (Software Engineer)

    \faThumbsODown  ~ \say{\textit{gRPC should be \textbf{avoided at any time}.}} (Architect)
\end{myquote}

{\renewcommand{\arraystretch}{1.5}
\begin{table*}[]
\caption{Security practices for \textbf{private microservices} and the survey responses (in \%). \textbf{AU}: Absolutely Useful, \textbf{U}: Useful, \textbf{NU}: Not Useful, \textbf{ANU}: Absolutely Useful, \textbf{IDK}: I Don't  Know, \textbf{MED}: Median, \textbf{AVG}: Average}
\label{group5}
\resizebox{\textwidth}{!}{
\begin{tabular}{ l p{10cm} | c | c c c c c | c c }
\hline
             &    \textbf{Security practices for private microservices}  & \textbf{Sources}                                                                                                                                          & \textbf{AU} & \textbf{U} & \textbf{NU} & \textbf{ANU} & \textbf{IDK} & \textbf{MED} & \textbf{AVG}  \\ \Xhline{2.3\arrayrulewidth}
\textbf{PP1} & Remote nodes (e.g., remote microservices) should not be able to even list/check for the existence of private microservices.                           & \cite{practicePP1_PP2_PI5}
&
\cellcolor[HTML]{B0FF99} 40.54        & 
\cellcolor[HTML]{CEFDBF} 27.03       & 
\cellcolor[HTML]{DFFFD6} 13.51        & 
\cellcolor[HTML]{FFFFFF} 0.00            & 
\cellcolor[HTML]{DBFFCE} 18.92           & 3.50               & 3.33             \\ 
\textbf{PP2} & Use “service grouping” to limit the visibility and callability (of both actions and events) of the private microservices in a group of microservices. & \cite{practicePP1_PP2_PI5}
&
\cellcolor[HTML]{C0FFAD} 32.43        & 
\cellcolor[HTML]{BDFFA6} 33.78       & 
\cellcolor[HTML]{EDFDE8} 6.76         & 
\cellcolor[HTML]{FFFFFF} 0.00            & 
\cellcolor[HTML]{CEFDBF} 27.03         & 3               & 3.35             \\ \hline
\end{tabular}}
\end{table*}}

\subsubsection{Private Microservices}\label{Private-Microservices}
Table \ref{group5} provides 2 practices to increase the security of private microservices. Private microservices are internal microservices in an organization that only a specific group of end-users or applications can access.

\faLightbulbO{} \textsc{\textbf{PP1.}} \textit{Remote nodes (e.g., remote microservices) should not be able to even list/check for the existence of private microservices.} If there are some private microservices in a microservices architecture, none of the remote nodes must be allowed to call private microservices' actions \cite{practicePP1_PP2_PI5}. In addition, they should not be allowed to check any information about private microservices directly (even the number of private microservices). In this scenario, private microservices can only contact internal microservices. 
50 out of 74 survey respondents (67.57\%) considered this practice \textit{absolutely useful} or \textit{useful}. We also did not receive \textit{absolutely not useful} for this practice. Two participants shared their feedback on this practice as follow:

\begin{myquote}

    \faThumbsOUp ~ \say{\textit{The requirement/responsibility of checking for the existence [of private microservices] may cause \textbf{tight coupling between services} which may cause availability issues and also potential “distributed monolith” because a service may not function if other services are unavailable.}} (Developer)
    
    
\end{myquote}

\faLightbulbO{} \textsc{\textbf{PP2.}} \textit{Use “service grouping” to limit the visibility and callability (of both actions and events) of the private microservices in a group of microservices.} More than 65\% of the survey participants (49 out of 74) acknowledged that the use of service grouping for private microservices is (\textit{absolutely}) \textit{useful}. This practice received the highest rate of \say{I don't know} among all practices (27.03\%). However, \textbf{PP2} did not receive \textit{absolutely not useful}, and only a few participants (6.76\%) rated it as \textit{not useful}.

An architect commented that service grouping is a beneficial method in microservices systems and is always required. A technical lead also detailed that using this practice can lead to preventing the awareness of systems and improving security.

\begin{myquote}
    \faThumbsOUp ~ \say{\textit{As a best practice from Microservices or Containerization point of view, \textbf{service grouping is always required} and beneficial.}} (Architect)
    
    \faThumbsOUp ~ \say{\textit{Ensuring visibility is restricted to the necessary services \textbf{is an essential part of zero trust}. It \textbf{prevents the awareness of systems, improving security}.}} (Technical Lead)
\end{myquote}

{\renewcommand{\arraystretch}{1.5}
\begin{table*}[]
\caption{Security practices for \textbf{database and environments} and the survey responses (in \%). \textbf{AU}: Absolutely Useful, \textbf{U}: Useful, \textbf{NU}: Not Useful, \textbf{ANU}: Absolutely Useful, \textbf{IDK}: I Don't  Know, \textbf{MED}: Median, \textbf{AVG}: Average}
\label{group6}
\resizebox{\textwidth}{!}{
\begin{tabular}{ l p{9.5cm} | c | c c c c c | c c }
\hline
             &     \textbf{Security practices for database and environments}      & \textbf{Sources}                                                                                                                                                                                                            & \textbf{AU} & \textbf{U} & \textbf{NU} & \textbf{ANU} & \textbf{IDK} & \textbf{MED} & \textbf{AVG} \\ \Xhline{2.3\arrayrulewidth}
\textbf{PD1} & Although security policies should be applied in both development and production environments, production environments need stronger security.                                                                                 &  
\cite{practicePI7andPD1}, \cite{practicePD1_2}, \cite{practicePD1_3}         
& 
\cellcolor[HTML]{7CFF54} 68.92        & 
\cellcolor[HTML]{D5FFC9} 20.27       & 
\cellcolor[HTML]{F3FFF1} 5.41         & 
\cellcolor[HTML]{F9FFF7} 2.70          & 
\cellcolor[HTML]{F9FFF7} 2.70          & 
4              & 3.6            \\ 
\textbf{PD2} & In a microservices architecture, databases should not be exposed to any unauthenticated request.                                                                                                                                  &       \cite{practicePD2}                     
& 
\cellcolor[HTML]{7AFF51} 71.63        & 
\cellcolor[HTML]{D5FFC9} 18.92       & 
\cellcolor[HTML]{F5FEF2} 4.05         & 
\cellcolor[HTML]{F9FFF7} 2.70          & 
\cellcolor[HTML]{F9FFF7} 2.70          & 4               & 3.64             \\ 
\textbf{PD3} & Suppose a microservice needs to validate some data against data from another microservice synchronously. In that case, it is recommended to combine both microservices and have only one microservice.                     &  \cite{practicePD3andPD4}      
& 
\cellcolor[HTML]{EBFFE4} 10.82         & 
\cellcolor[HTML]{D4FFC7} 24.32       & 
\cellcolor[HTML]{ABFF91} 40.54        & 
\cellcolor[HTML]{E5FFDD} 13.51         & 
\cellcolor[HTML]{E5FFDD} 10.81         & 2              & 2.36             \\ 
\textbf{PD4} & Suppose a microservice needs to validate some data against data from another microservice synchronously. In that case, the first microservice should replicate data from the second microservice in its own database with an eventual consistency syncing system. &  \cite{practicePD3andPD4}
&
\cellcolor[HTML]{DFFFD6} 18.92        & 
\cellcolor[HTML]{C9FFAB} 35.14       & 
\cellcolor[HTML]{CBFFBA} 24.32          & 
\cellcolor[HTML]{F5FEF2} 4.05          & 
\cellcolor[HTML]{DBFFCE} 17.57           & 3               & 2.84              \\ \hline
\end{tabular}}
\end{table*}}

\subsubsection{Database and Environments}\label{Database-and-Environments}
Four practices are categorized into the database and environments group and are shown in Table \ref{group6}. They focus on security concerns that databases and production environments may raise in microservices systems.

\faLightbulbO{} \textsc{\textbf{PD1.}} \textit{Although security policies should be applied in both development and production environments, production environments need stronger security.} The majority of the survey participants (89.19\%) accepted that production environments need more security policies than development environments. Among these participants, more than 65\% rated it as \textit{absolutely useful}. Less than 10\% disagreed with the usefulness of this practice. In the following, we received some comments that support or refute/question \textbf{PD1}:

\begin{myquote}
    \faThumbsOUp ~ \say{\textit{Because different kinds of clients with different \textbf{unknown intentions will access} the system.}} (Developer)
\end{myquote}

\begin{myquote}

    \faThumbsODown ~ \say{\textit{It is best to keep development and production environments \textbf{as similar as possible} to avoid \textbf{surprise issues}.}} (Software Engineer)
    
    \faThumbsODown ~ \say{\textit{Best practice would \textbf{enable the same} in development environment.}} (DevOps Engineer)
    
    \faThumbsODown ~ \say{\textit{Depending on the information, the development environment needs to have a \textbf{security level equivalent to} the production one.}} (Software Engineer)

    \faThumbsODown ~ \say{\textit{Security should be \textbf{repeated across all environments}.}} (Software Engineer)
\end{myquote}

From the comments, some participants believed that the production environment should enable security policies more than the development one because various types of clients with unknown intentions can use the microservices system in the production environment. Conversely, other participants believed that development and production environments should have equal security policies to make them identical as much as possible.

\faLightbulbO{} \textsc{\textbf{PD2.}} \textit{In a microservices architecture, databases should not be exposed to any unauthenticated request.}
The clients send various requests to resources in a microservices architecture. Since the resources always contain important information, it is extremely recommended that no resources accept the requests which are not authenticated \cite{practicePD2}. We received a high rate of usefulness (\textit{absolutely useful} or \textit{useful}) for \textbf{PD2} (90.55\%) which more than 70\% of them were \textit{absolutely useful}. A few comments which support or refute this practice are shown below:

\begin{myquote}
    \faThumbsOUp ~ \say{\textit{\textbf{Unauthenticated requests should not} be enabled on database.}} (Technical Lead)
    
    \faThumbsOUp ~ \say{\textit{Needs to have its \textbf{default secure authentication}, as well as its own private VPN network.}} (Software Engineer)
    
    \faThumbsODown ~ \say{\textit{There are some cases that don't require authentication and it requires to do database operation, e.g., \textbf{scheduler to clean up log table}.}} (Architect)
\end{myquote}

\faLightbulbO{} \textsc{\textbf{PD3.}} \textit{Suppose a microservice needs to validate some data against data from another microservice synchronously. In that case, it is recommended to combine both microservices and have only one microservice.} Almost 65\% of the survey participants disagreed with the usefulness of the practice or indicated that they had no idea about this practice (see Table \ref{group6}). 
Still, 35.14\% of our survey participants chose \textbf{PD3} as (\textit{absolutely}) \textit{useful}. 

The main reason stated this practice is not useful is that it can be against the Single Responsibility principle and increase the size of microservices.

\begin{myquote}

 \faThumbsODown ~ \say{\textit{[In] some \textbf{exceptional cases,} this item might be useful but in most of the cases if we follow this we \textbf{will end up a few huge services} instead of a real microservices system.}} (Software Engineer)
    
    \faThumbsODown ~ \say{\textit{To handle the synchronization scenario, I don't think that it is a good idea to \textbf{break the Single Responsibility principle} rather eventual consistency needs to follow.}} (Architect)
\end{myquote}

\faLightbulbO{} \textsc{\textbf{PD4.}} \textit{Suppose a microservice needs to validate some data against data from another microservice synchronously. In that case, the first microservice should replicate data from the second microservice in its own database with an eventual consistency syncing system.} Nearly 55\% of the survey participants agreed with this practice, and less than 30\% of our survey participants voted \textbf{PD4} as (\textit{absolutely}) \textit{not useful}. Two positive comments on this practice are:

\begin{myquote}

\faThumbsOUp ~ \say{\textit{The recommendation in case of a sync would be to actually \textbf{keep a copy of the data} in the two services, remembering the \textbf{Consistency-Availability-Partition tolerance (CAP) theorem}.}} (Software Engineer)

\faThumbsOUp ~ \say{\textit{One of the problems working with microservices is \textbf{data replication}. This should not be a problem if you guarantee that your data will always be updated. You can \textbf{use saga pattern or event out box pattern} for this.}} (Software Engineer)

\end{myquote}

\section{Recommendations}\label{recommendations}

In this section, we present concrete and actionable recommendations for microservices practitioners and researchers based on our reflections on the findings. 

\subsection{Recommendation for Practitioners}\label{sec:recompractitioners}

\textit{\textbf{The most useful practices}}. All 28 practices, except for practice \textbf{PD3}, tend to have the median Likert score of 3, 3.5 or 4, indicating that the vast majority of the survey participants affirmed these practices are \textit{useful} or \textit{absolutely useful}. At the same time, we acknowledge that software organizations and practitioners may not be willing to or cannot adopt all 28 security practices (e.g., lack of enough resources). Hence, we highlight the 8 most important security practices, including \textbf{PA4}, \textbf{PI3}, \textbf{PI4}, \textbf{PI5}, \textbf{PI6}, \textbf{PM2}, \textbf{PD1}, and \textbf{PD2}, with a median Likert score of 4 (\textit{absolutely useful}) and encourage microservices practitioners to adopt these security practices to ensure the desired level of security in microservices systems.

\textbf{PA4} from the “authorization and authentication” group emphasizes using API Gateway to handle authorizing or routing microservices in large-scale microservices systems. Four out of these 8 highly accepted practices, including \textbf{PI3}, \textbf{PI4}, \textbf{PI5}, and \textbf{PI6}, come from the “internal and external microservices” group. This can (partially) show the importance of the network of microservices in terms of internal and external domains and the level of security policies and practices that should be considered. \textbf{PM2} from the ``microservices communications'' group shows that using security protocols for the communication between microservices and databases is important.
\textbf{PD1} and \textbf{PD2} focus on security concerns in production environments and databases. \textbf{PD1} argues that microservices systems need more security policies when executing in production environments compared to development environments. \textbf{PD2} informs microservices practitioners that microservices' databases should not accept any types of unauthenticated requests.

\textit{\textbf{Several factors still matter while using security practices}}. Although the survey participants considered almost all identified security practices useful, we do not claim that these 28 practices are the best options for all contexts and domains. We assert that practitioners should carefully consider different context-sensitive factors and trade-offs when using each of these practices. According to the responses of our participants, such factors and trade-offs are enormous, ranging from design context to user experience, from required security skills and expertise to infrastructure resources. For example, it is important to consider if the given system or service is public or private. What are the impacts of the security practices on other quality attributes (e.g., performance)? How sensitive is the data? It is also important to consider to what extent a security practice may impact the user experience. Cost and complexity associated with some practices were mentioned by several respondents as other important factors. A Technical Lead with more than 5 years of experience in microservices systems summarized it as: \say{\textit{All scenarios [practices] are useful, depending on system requirements. Security is an essential overhead; however, it is an overhead (with speed and complexity). So be sensible, don’t over-engineer it. Equally, make sure sensitive data cannot be leaked - use private networks where possible.}}


\subsection{Recommendations for Researchers}

\textit{\textbf{Study how the identified practices are used in different microservices systems in different domains and contexts.}} In Section \ref{sec:recompractitioners}, we discussed that although most of the survey participants affirmed the usefulness of the identified security practices, the successful implementation of these practices depends on too many factors and trade-offs. In this study, we have tried, to some extent, to show in which circumstances some practices are useful (e.g., \textbf{PT3}, \textbf{PI7}) or their impacts on other quality attributes (e.g., \textbf{PA1}, \textbf{PA6}). However, it is out of the scope of our study to explore and discuss all factors and trade-offs. More efforts should be allocated to investigate the short-term and long-term impacts of each of the identified practices, their impacts on a specific type of microservices systems, e.g., IoT microservices systems, and their associated costs and overhead.


\textit{\textbf{Study why some practices are controversial.}}  There are still some practices that were slightly controversial (e.g., \textbf{PA1}, \textbf{PA2}, \textbf{PA3}, \textbf{PT4}, \textbf{PI1}, \textbf{PI2}, \textbf{PI3} and \textbf{PD1}). For example, \textbf{PT4} suggests decoding JWT at the microservices level instead of the API Gateway level (i.e., more than 60\% agreed with this practice). At the same time, 25.68\% of the practitioners still thought that decoding JWT at the API Gateway level is better than at the microservices level. Also, the survey respondents tended to disagree with a few security practices (e.g., \textbf{PD3} and \textbf{PD4}) that we found from GitHub and Stack Overflow. There might be several reasons behind controversial practices. As we discussed before, several factors (e.g., the design context and user experience) may impact the opinion of practitioners on using or not using security practices in a software system. Given that the security practices were detected in open-source projects and the survey respondents came from both open-source projects and industrial projects, they might have had a different experience in implementing these practices. Finally, the security practices were provided to the survey respondents in 1-2 sentences, which might not be the best way to describe all aspects of some practices. This can be another reason to cause some controversial practices. Thereby, we argue that an important research direction in future could be exploring the reasons behind controversial practices in securing microservices systems. 


\textit{\textbf{Pay attention to security practices in other or less explored aspects of microservices systems.}}
In this study, we only looked at 2 developer discussion platforms (10 microservices systems from GitHub and 306 posts from Stack Overflow) to identify 28 security practices categorized into 6 groups. We do emphasise that these 28 practices are only a subset of available and required security practices for microservices systems. We believe that many security practices were not discussed or could not be found in our data sources. For example, we were not able to find any concrete practices and guidelines regarding security audits, how to safely recover from security failures and how to secure data in microservices systems. Potential security risks associated with tools and technologies used in microservices system development and deployment also play an important role in achieving secure microservices systems. For example, container images generated by third parties (e.g., Docker) may be associated with several security risks. However, our analysis of open-source projects hosted on GitHub and Stack Overflow posts did not find any concrete practices on how to address the security risks. While some other works (e.g., \cite{pereira2021security, hannousse2021securing}) have recently examined (gray) literature to understand security in microservices systems, we encourage researchers to mine other sources, such as other developer discussion platforms (e.g., Reddit\footnote{\url{https://www.reddit.com}}) to identify more practices. These new practices can either complement our security practices in a given group (e.g., the “database and environments” group) or be new categories of security practices (e.g., security practices for containerized microservices systems).

\section{Threats to Validity}\label{Threats-to-Validity}
In this section, we summarize potential threats to the validity of this study and the strategies that we used to mitigate these threats \cite{wohlin2012experimentation}.

\subsection{External Validity}
Three threats might limit the generalizability of our findings. First, we identified the 28 security practices from only 2 data sources: GitHub and Stack Overflow. Although these platforms are the most popular online platforms among different types of software practitioners to share and discuss software development challenges, knowledge, and solutions, they do not represent all views of software practitioners. Second, we chose 10 open-source microservices systems on GitHub, which is only one popular software repository. These projects vary in terms of domain, number of contributors, size, etc. Despite this fact, we cannot claim that these projects are representative of all types of microservices systems (e.g., IoT microservices systems) and all the OSS repositories. Our validation study did not receive many responses (i.e., 63), and not all the respondents answered the open-ended questions in the survey.
This threat was, to some extent, reduced as software practitioners with different characteristics (e.g., possessing different roles and working in organizations with diverse domains) completed the survey.

\subsection{Internal Validity}
Identifying security practices from Stack Overflow and GitHub might be subjective and error-prone. We adopted several strategies to reduce this issue. First, several analysts and validators were involved in this process. Three analysts participated in the pilot phase and the main phase of the data analysis (see Section \ref{subsec:identificationofsecuritypractices}). Three other authors with extensive experience in security in microservices systems reviewed and validated the identified security practices and suggested some feedback. Finally, we deployed a pilot survey to seek practitioners' feedback on the identified security practices. This helped us remove 4 practices and improved the wording of some of the security practices.


The validation survey tried to recruit practitioners with experience in securing microservices systems. As discussed in Section \ref{participants}, we used different recruitment approaches for this purpose. One of the approaches was to carefully check the profiles of many practitioners on their websites, Slack, LinkedIn, etc. This approach might have led to two threats. We might have mainly recruited practitioners who successfully applied security in microservices system development and ignored the entire microservices practitioners (e.g., unsuccessful practitioners in applying security in microservices systems). This bias is referred to as survivorship bias \cite{brown1992survivorship}. On the other hand, still practitioners with poor knowledge of the MSA style and security may have participated in the survey, which can be a concern for the validity of the survey.

We employed some solutions to (partially) mitigate these threats. Our survey was not filled out by only a few specific roles and did not recruit the respondents using only one recruitment approach. Practitioners with different roles who have worked on various open-source and industrial projects, such as developers, software engineers, DevOps engineers, requirements engineers, and architects, completed the validation survey. We also added the \say{I Don't Know} option in the survey questions to minimize the concern of lack of knowledge on some identified practices. We also asked the survey respondents to comment on why they rated a given practice \say{Useful} or \say{Not Useful}. The detailed comments from the survey respondents increased our confidence that the vast majority of them had the right experience and expertise. We invited the contributors of the 10 open-source projects from which the some of best practices were extracted to complete the survey. We acknowledge that the survey responses coming from the contributors of the 10 open-source projects might have provided a biased assessment of the security practices. However, given that we used different recruitment approaches, the percentage of survey responses coming from the contributors of the 10 open-source projects should not be high.


\subsection{Construct Validity}
Our decision to use DeepM1 introduced in \cite{rezaei2021automated} to extract security paragraphs from cortex, spinnaker, and jaeger projects might have introduced threats. Although DeepM1 has a good performance in detecting security paragraphs from GitHub issues and Stack Overflow posts concerning security in microservices systems, we acknowledge that we might have missed some important security information from these projects. Furthermore, we defined security points as an issue or post with equal to or more than 5 security paragraphs. We admit that some issues or posts with less than 5 security paragraphs may still contain important microservices security practices. In this study, we only used the validation survey to evaluate the usefulness of the identified practices. Other research methods such as case studies could also be used to indicate all positive and negative aspects of the identified security practices.

\subsection{Reliability}
There is a potential threat that other researchers replicate our study and generate different results. Our first approach to alleviate this threat was to provide a detailed explanation of our research method (e.g., the survey design), enabling other researchers to replicate our study. Furthermore, we created a replication package \cite{onlinedataset}, including the 861 security points used to identify security practices and the encoded survey responses, allowing other researchers and practitioners to validate our findings.

\section{Conclusions and Future Work}\label{ConclusionsandFutureWork}

This study identified 28 security practices for securing microservices systems through manually examining 861 microservices security points. These 861 microservices security points include 543 GitHub issues, 9 official documents, and 3 wiki pages from 10 open-source microservices systems, and 306 Stack Overflow posts concerning security in microservices systems. These 28 security practices are categorized into 6 groups: \textit{Authorization and Authentication}, \textit{Token and Credentials}, \textit{Internal and External Microservices}, \textit{Microservices Communications}, \textit{Private Microservices}, and \textit{Database and Environments}. Through an online survey completed by 74 microservices practitioners, we have shown that the majority of the respondents rated these 28 practices useful for industrial usage.

In the future, we plan to extend our catalog of security practices by exploring more resources (e.g., interviews) to identify more security practices, in particular, in less explored areas of microservices systems. We also aim to investigate the positive and negative impacts of the identified security practices in different types of microservices systems.

\section*{\textbf{Acknowledgements}}

This work is funded by the National Natural Science Foundation of China (NSFC) with Grant No. 62172311 and the Special Fund of Hubei Luojia Laboratory. 

\balance

\bibliographystyle{elsarticle-num}

\bibliography{JSS2020}

\end{document}